\documentclass[aps,prl,twocolumn,groupedaddress]{revtex4-1}
\usepackage{graphicx}
\usepackage{amsmath}
\usepackage{amssymb}
\usepackage{braket}
\usepackage{comment}
\usepackage{color}
\usepackage{bm}
\newtheorem{thm}{Theorem}
\newtheorem{remark}{Claim}

\begin{document}

\title{Chemical Thermodynamics for Growing Systems}

\author{Yuki Sughiyama}
\thanks{These two authors contributed equally.}
\affiliation{Institute of Industrial Science, The University of Tokyo, 4-6-1, Komaba, Meguro-ku, Tokyo 153-8505 Japan}
\author{Atsushi Kamimura}
\thanks{These two authors contributed equally.}
\affiliation{Institute of Industrial Science, The University of Tokyo, 4-6-1, Komaba, Meguro-ku, Tokyo 153-8505 Japan}
\author{Dimitri Loutchko}
\affiliation{Institute of Industrial Science, The University of Tokyo, 4-6-1, Komaba, Meguro-ku, Tokyo 153-8505 Japan}
\author{Tetsuya J. Kobayashi}
\email[E-mail me at:]{tetsuya@mail.crmind.net}
\altaffiliation[Also at ]{Universal Biology Institute, The University of Tokyo, 7-3-1, Hongo, Bunkyo-ku, 113-8654, Japan.}
\affiliation{Institute of Industrial Science, The University of Tokyo, 4-6-1, Komaba, Meguro-ku, Tokyo 153-8505 Japan}
\date{\today}
\begin{abstract}
We consider growing open chemical reaction systems (CRSs), in which autocatalytic chemical reactions are encapsulated in a finite volume and its size can change in conjunction with the reactions.
The thermodynamics of growing CRSs is indispensable for understanding biological cells and designing protocells by clarifying the physical conditions and costs for their growing states.
In this work, we establish a thermodynamic theory of growing CRSs by extending the Hessian geometric structure of non-growing CRSs.
The theory provides the environmental conditions to determine the fate of the growing CRSs; growth, shrinking or equilibration.
We also identify thermodynamic constraints; one to restrict the possible states of the growing CRSs and the other to further limit the region where a nonequilibrium steady growing state can exist. 
Moreover, we evaluate the entropy production rate in the steady growing state. 
The growing nonequilibrium state has its origin in the extensivity of thermodynamics, which is different from the conventional nonequilibrium states with constant volume.
These results are derived from general thermodynamic considerations without assuming any specific thermodynamic potentials or reaction kinetics; i.e., they are obtained based solely on the second law of thermodynamics. 
\end{abstract}
\maketitle
\section{I. Introduction}
\label{SecI}
Self-replication is a hallmark of living systems by which they are differentiated from nonliving ones.
Since von Neumann’s formulation of self-reproducing automata \cite{vonNeumann,freitas01}, the physical and chemical basis of self-replication has been pursued theoretically and experimentally in order to understand and synthesize living systems \cite{andrieux01,pekar01,dourado01,thomas01,liu01,furusawa01,lin01,maitra01,pandey01,pandey02,roy01,joyce01,protocell01,noireaux01,kurihara01,kurihara02,kita01,ichihashi01,protocell01,protocell02,protocell03,segre01,protocell04,himeoka01}. 
Of the various components necessary for self-replication, autocatalytic reaction cycles, thought of as the driving engine, form a  central part \cite{unterberger01,autocatalytic_core,hypercycle,barenholz01,kauffman01,jain01,steel01}.
However, the presence of cycles is not sufficient for self-replication.
Because the cycles should be confined in an encapsulating volume which defines the replication unit, the size of the volume should also grow in accordance with the production of chemicals by the cycles.

In spite of the active investigation of autocatalytic reaction cycles in the last decades \cite{unterberger01,autocatalytic_core,hypercycle,barenholz01,kauffman01,jain01,steel01}, the growth of volume and its coupling with the autocatalytic cycles have not been thoroughly investigated so far.
Although the recent rediscovery of growth laws of bacteria \cite{growthlaw01} initiated a surge of new coarse-grained autocatalytic models \cite{growthlaw02,growthlaw03,growthlaw04,maitra01,reuveni01,reuveni02,reuveni03,pandey01,pandey02,roy01}, the volume growth in these models is considered only heuristically \cite{lin01,muller02,muller01,Daan,Hidde}, e.g., by representing it with a linear function of chemicals in it. 

In the light of chemical thermodynamics, the change in volume and the influx and outflux of chemicals driven by the cycles are mutually dependent and should be thermodynamically consistent.
This interdependence of reactions and volume inevitably constrain their possible states and dynamics. 
In addition, the cycles themselves may not always proceed in the forward direction to grow, depending on the environmental conditions. If it proceeds in the reverse direction, it can result in shrinking. 
It is nontrivial under what thermodynamic conditions a coherent forward cycle dynamics and volume growth can be achieved. 
Moreover, a steady cycling and growth should accompany the thermodynamic cost. 
However, we lack a theoretical basis to address these fundamental problems of growing systems.

In this work, we establish the thermodynamics for growing systems.
The difficulty in developing it lies in the fact that the change in the volume affects all reactions in it. 
In the conventional theory of chemical reactions, reaction fluxes are described as functions of densities of chemicals (concentrations) \cite{07,06,05,04,m6,m5,m4,m1}, which presumes a constant volume. 
However, if the volume changes, the densities can change even though the numbers of chemicals remain unchanged. 
Hence, it is necessary to return to a thermodynamic formulation in which the numbers of chemicals and the volume are treated separately.
In other words, we have to explicitly take account of the extensivity of thermodynamic functions, which is scaled out when the densities alone are considered.
Nevertheless, we should also retain the density representation and its dual representation by the chemical potentials to appropriately characterize steady growing states and the conditions imposed by the intensive variables of the environment. 

We clarify this entangled relation among the triad of chemical numbers, densities and potentials by identifying the geometric structure they form.
This structure is built on the recently discovered Hessian geometric structure between chemical densities and potentials in a constant volume \cite{sughiyama01,kobayashi01} by additionally introducing the space of the numbers of chemicals.
Based on the second law of thermodynamics, our theory classifies the thermodynamic conditions under which the system grows, shrinks or equilibrates. 
It also reveals the region in which the chemical density is constrained to a steady growth.
Furthermore, it enables us to evaluate the entropy production rate, i.e., the physical cost of the steady growth. 
Our nonequilibrium system with volume growth has its origin in the extensivity of thermodynamics, which is different from the conventional nonequilibrium systems with constant volume \cite{07,06,05,04,m6,m5,m4,m1}. 

We emphasize that our derivation is performed based on a purely thermodynamic argument \cite{sughiyama01,kobayashi01,thermo1,thermo2}. 
As a result, it does not depend on any particular form of thermodynamic potentials or reaction kinetics \cite{reply2}.
Thus, our theory is widely applicable and contributes to understanding the origins of life and constructing protocells \cite{protocell01,protocell02,protocell03,protocell04,joyce01,noireaux01,kurihara01,kurihara02,kita01,ichihashi01,segre01,himeoka01} as well as seeking the universal laws of biological cells \cite{growthlaw01,growthlaw02,growthlaw03,growthlaw04,reuveni01,reuveni02,reuveni03,roy01,maitra01,pandey01,pandey02}.
Moreover, a more realistic thermodynamic cell model may be constructed by integrating various other components such as active transport, responsive kinetics of the membrane, metabolism, etc.

This paper is organized as follows. 
We devote Sec. II to outline our main results without showing the details of their derivation. 
From Sec. III onward, we start with the derivation of our main results. 
In Sec. III, we analyze the behavior of the total entropy function with respect to time for chemical reaction dynamics.
We devote Sec. IV to the preparation for the geometric structure of growing systems. 
In Sec. V, we classify the environmental conditions to determine the fate of the system based on the form of the total entropy function.
In Sec. VI, we consider the steady growing state and evaluate the entropy production rate in this state. 
We illustrate our theory in Sec. VII for the ideal gas as a specific example of thermodynamic potentials. 
In Sec. VIII, we numerically verify our theory by considering a specific example of a chemical reaction system composed of the ideal gas and obeying mass action kinetics.  
Finally, we summarize our work with further discussions in Sec. IX.

\section{II. Outline of the main results}
\label{SecII}

\subsection{A. Thermodynamic setup}
\begin{figure}
    \centering
    \includegraphics[width=0.5\textwidth, clip]{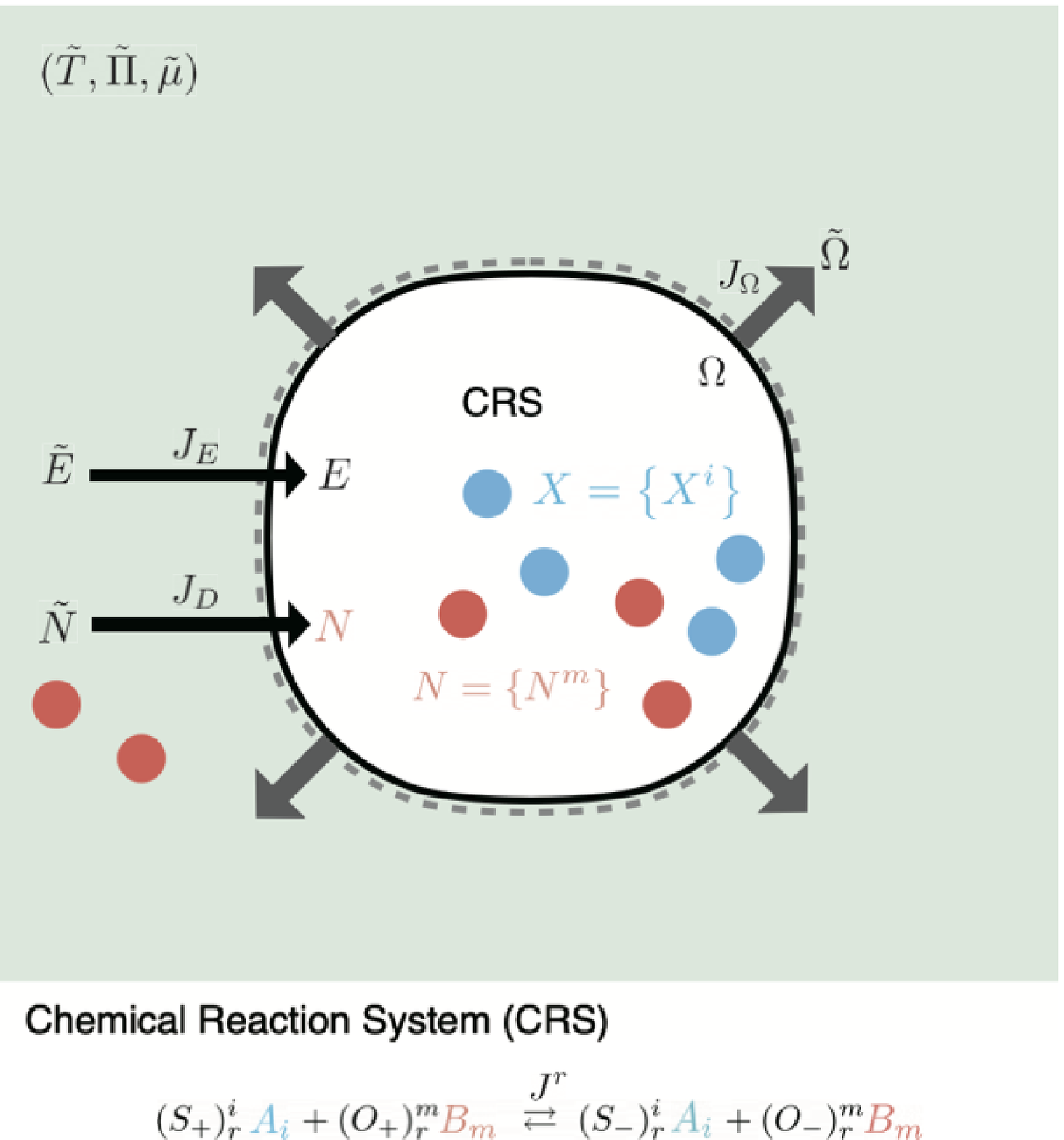}
    \caption{Diagrammatic representation of open CRSs. The chemical reactions occur with the reaction fluxes $J(t) = \{J^r(t)\}$, the $r$th reaction of which is represented as the chemical equation at the bottom. Here, $A = \{ A_i \}$ are the labels of the confined chemicals, and $B = \{ B_m\}$ are the ones of the open chemicals which can move across the membrane with the diffusion fluxes $J_D(t)=\{J_D^m(t)\}$. 
    The numbers of the confined and open chemicals in the system are denoted by $X = \{X^i\}$ and $N = \{N^m\}$, respectively. 
    Also, $(S_+)^i_r$ and $(O_+)^m_r$ denote stoichiometric coefficients of the reactants in $r$th reaction, whereas $(S_-)^i_r$ and $(O_-)^m_r$ are the ones of the products. The stoichiometric matrices are given as $S^i_r = (S_-)^i_r - (S_+)^i_r$ and $O^m_r = (O_-)^m_r - (O_+)^m_r$.
    For theoretical simplicity, we ignore the tension of the membrane and assume that it never bursts. }
    \label{fig:system}
\end{figure}

Let us start with the presentation of the setting of the system (FIG. \ref{fig:system}). 
Consider a growing open chemical reaction system (CRS) surrounded by a reservoir.
We assume that the system is always in a well-mixed state (a local equilibrium state), and therefore we can completely describe it by extensive variables $(E,\Omega,N,X)$. 
Here, $E$ and $\Omega$ represent the internal energy and the volume; $N=\left\{N^{m}\right\}$ denotes the number of  chemicals that can move across the membrane between the system and the reservoir called open chemicals; 
meanwhile, $X=\left\{X^{i}\right\}$ is the number of chemicals confined within the system; the indices $m$ and $i$ run from $m=1$ to $\mathcal{N}_N$ and from $i=1$ to $\mathcal{N}_X$, respectively, 
where $\mathcal{N}_N$ and $\mathcal{N}_X$ are the numbers of species of the open and confined chemicals.     
The reservoir is characterized by intensive variables $(\tilde{T},\tilde{\Pi},\tilde{\mu})$, where $\tilde{T}$ and $\tilde{\Pi}$ are the temperature and the pressure; $\tilde{\mu}=\left\{\tilde{\mu}_{m}\right\}$ is the chemical potential corresponding to the open chemicals. 
Also, we denote the corresponding extensive variables by $(\tilde{E},\tilde{\Omega},\tilde{N})$. 

In thermodynamics, the entropy function is defined on $(E,\Omega,N,X)$ as a concave, smooth and homogeneous function $\Sigma\left[E,\Omega,N,X\right]$. 
We write the entropy function for the reservoir as $\tilde{\Sigma}_{\tilde{T},\tilde{\Pi},\tilde{\mu}}[\tilde{E},\tilde{\Omega},\tilde{N}]$, and therefore the total entropy can be expressed as 
\begin{equation}
\Sigma^{\mathrm{t}\mathrm{o}\mathrm{t}}=\Sigma\left[E,\Omega,N,X\right]+\tilde{\Sigma}_{\tilde{T},\tilde{\Pi},\tilde{\mu}}\left[\tilde{E},\tilde{\Omega},\tilde{N}\right],\label{enttot}
\end{equation} 
where we use the additivity of the entropy. Furthermore, due to the homogeneity of the entropy function for the system, without loss of generality, we can write it as 
\begin{equation}
\Sigma\left[E,\Omega,N,X\right]=\Omega\sigma\left[\epsilon,n,x\right],
\end{equation}
where $\sigma\left[\epsilon,n,x\right]$ is the entropy density and $\left(\epsilon,n,x\right):=(E/\Omega,N/\Omega,X/\Omega)$. 
Since this work only treats a situation without phase transitions, we assume that $\sigma\left[\epsilon,n,x\right]$ is strictly concave.

Next, we define the dynamics for the system as 
\begin{eqnarray}
\displaystyle \nonumber&&\frac{dE}{dt}=J_{E}\left(t\right),\mbox{  }\frac{d\Omega}{dt}=J_{\Omega}\left(t\right),\\
&&\displaystyle \frac{dN^{m}}{dt}=O_{r}^{m}J^{r}\left(t\right)+J_{D}^{m}\left(t\right),\mbox{  }\frac{dX^{i}}{dt}=S_{r}^{i}J^{r}\left(t\right),\label{SDynamics}
\end{eqnarray}
where $J_{E}\left(t\right),\ J_{\Omega}\left(t\right),\ J_{D}\left(t\right)=\left\{J_{D}^{m}\left(t\right)\right\}$ and $J\left(t\right)=\left\{J^{r}\left(t\right)\right\}$ represent the energy, the volume, the chemical diffusion and the chemical reaction fluxes, respectively; 
$S=\left\{S_{r}^{i}\right\}$ and $O=\left\{O_{r}^{m}\right\}$ denote stoichiometric matrices for the confined and the open chemicals (see FIG. \ref{fig:system}). 
The index $r$ runs from $r=1$ to $\mathcal{N}_R$, where $\mathcal{N}_R$ is the number of reactions. 
Also, in Eq. (\ref{SDynamics}), we employed Einstein's summation convention for notational simplicity.
The dynamics of the reservoir is given as
\begin{equation}
\displaystyle \frac{d\tilde{E}}{dt}=-J_{E}\left(t\right),\mbox{  }\frac{d\tilde{\Omega}}{dt}=-J_{\Omega}\left(t\right),\mbox{  }\frac{d\tilde{N}^{m}}{dt}=-J_{D}^{m}\left(t\right).\label{RDynamics}
\end{equation}

In this work, we assume that the time scale of the reactions is much slower than that of the others (that is, $J_{E}\left(t\right),J_{\Omega}\left(t\right),J_{D}(t) \gg J\left(t\right)$). 
Therefore, our dynamics is effectively governed only by the reaction flux $J(t)$ \textbf{(see Sec. III for details)}.
It means that we focus on the simplest thermodynamic setting in which the size of the volume is thermodynamically determined (see Eq. (\ref{revision3})). Thus, the active transport of material and responsive membrane kinetics are ignored for simplicity.
In addition, we assume the regularity of the stoichiometric matrix $S$ for the confined chemicals, i.e., $\mathcal{N}_X=\mathcal{N}_R=\mathrm{Rank}[S]$. 
This regularity was recently employed to identify minimal motifs of autocatalytic cycles, which were proposed in \cite{autocatalytic_core} (\textbf{see Appendix A for details}). 
We note that the regularity of $S$ is just a sufficient condition of the minimal motifs. 
Thus, our theory based only on the regularity of $S$ can be applied to a wider class of autocatalytic cycles than the minimal motifs.

\subsection{B. Thermodynamic potentials, duality, and total entropy characterizing the growing systems}
With the above setup, we obtain a conjugate pair of thermodynamic potentials, $\varphi(x)$ and $\varphi^*(y)$, which play pivotal roles in our theory.  
The partial grand potential density $\varphi(x) = \varphi[ \tilde{T}, \tilde{\mu}; x ]$ is defined as 
\begin{equation}
    \displaystyle \varphi\left[\tilde{T},\tilde{\mu};x\right]:=\min_{\epsilon,n}\left\{\epsilon-\tilde{T}\sigma\left[\epsilon,n,x\right]-\tilde{\mu}_{m}n^{m}\right\},
\end{equation}
\textbf{(see Sec. IV for details)}.
The function $\varphi^*(y) = \varphi^{*}[\tilde{T}, \tilde{\mu}; y]$ is the full grand potential density obtained by the Legendre transformation of $\varphi(x)$ as
\begin{equation}
\displaystyle \varphi^{*}\left[\tilde{T}, \tilde{\mu}; y\right]:=\max_{x}\left\{ y_{i}x^{i}-\varphi(x)\right\}.
\end{equation}
In conventional chemical thermodynamics with a constant volume, $\varphi(x)$ and $\varphi^*(y)$ characterize the system completely.
They also work as the dual convex functions inducing the Hessian geometric structure of chemical thermodynamics \cite{sughiyama01,kobayashi01}.
Because of the one-to-one correspondence of the Legendre transformation induced by $\varphi(x)$ and $\varphi^{*}(y)$, we can equivalently specify a state of the system either by the density $x$ or by its Legendre transform $y=\partial \varphi(x)$.
The thermodynamic interpretation of $y$ is the corresponding chemical potential to $x$. 
This dualistic representation is central to our theory.
In addition, $\varphi^*(y)$ can be interpreted as the pressure of the system at the state $y$ whose corresponding density is $x=\partial \varphi^{*}(y)$.

\begin{figure}
    \centering
    \includegraphics[width=0.5\textwidth, clip]{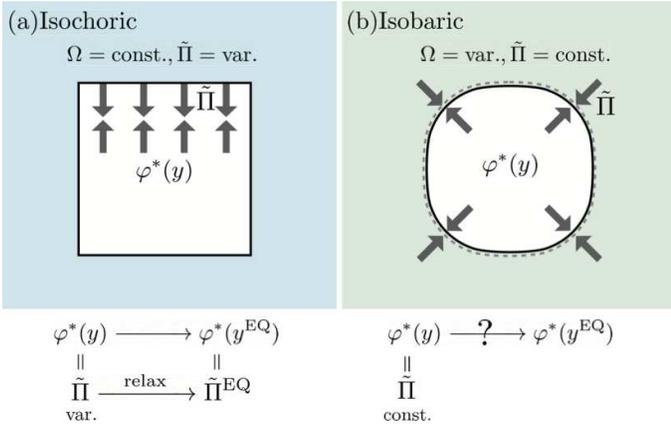}
    \caption{Diagrammatic representation of (a) isochoric and (b) isobaric situations.
    (a) In the isochoric case, the external pressure $\tilde{\Pi}$ varies to keep the volume $\Omega$ constant. The internal pressure $\varphi^{*}(y)$, which always balances with $\tilde{\Pi}$, can converge to the chemical equilibrium pressure $\varphi^{*}(y^\mathrm{EQ})=\tilde{\Pi}^\mathrm{EQ}$. 
    (b) In the isobaric case, the volume $\Omega$ varies to keep the internal pressure $\varphi^{*}(y)$ always equal to the constant external pressure $\tilde{\Pi}$. Consequently, the internal pressure $\varphi^{*}(y)=\tilde{\Pi}$ may not balance with the chemical equilibrium pressure $\varphi^{*}(y^\mathrm{EQ})$, which is specified by the chemical potentials $\tilde{\mu}$ in the reservoir.
    This imbalance drives growing or shrinking of the volume.}
    \label{fig:introbyD}
\end{figure}

If the volume is fixed, the internal pressure $\varphi^{*}(y)$ always balances with the external pressure $\tilde{\Pi}$ incurred by the boundary to keep the volume $\tilde{\Omega}$ constant (see FIG. \ref{fig:introbyD}(a)).
Furthermore, the internal pressure $\varphi^{*}(y)=\tilde{\Pi}$ converges to the pressure $\varphi^{*}(y^\mathrm{EQ})=\tilde{\Pi}^\mathrm{EQ}$ at the chemical equilibrium state $y^\mathrm{EQ}$. 
The state $y^\mathrm{EQ}$ is given by the solution to the simultaneous equations: 
\begin{equation}
y_{i}^{\mathrm{EQ}}S_{r}^{i}+\tilde{\mu}_{m}O_{r}^{m}=0,
\label{revision1}
\end{equation}
which describe the balances of chemical potentials between reactants and products at the chemical equilibrium \cite{sughiyama01}.
%The state $y^\mathrm{EQ}$ is determined by the chemical potentials $\tilde{\mu}$ in the reservoir as 
Since $S$ is regular, Eq. (\ref{revision1}) has the unique solution:
\begin{equation}
y_i^{\mathrm{E}\mathrm{Q}} = -\tilde{\mu}_m O_r^m \left(S^{-1}\right)^r_i,
\label{revision2}
\end{equation}
where $S^{-1}$ is the inverse of the stoichiometric matrix $S$ \cite{nr1}.
In the density representation, the system converges to the chemical equilibrium state $x_{\mathrm{E}\mathrm{Q}}=\partial \varphi^{*}(y^{\mathrm{E}\mathrm{Q}})$. 

By contrast, in growing systems under isobaric conditions, the volume can change. 
Due to the fast time scale of the volume flux $J_\Omega(t)$, the internal pressure $\varphi^{*}(y)$ is fixed by the external (reservoir) one $\tilde{\Pi}$ (see FIG. \ref{fig:introbyD}(b)). 
As a result, the volume at $X$ is variationally determined as
\begin{equation}
\displaystyle \Omega\left(X\right)=\arg\min_{\Omega}\left\{\Omega\varphi\left(\frac{X}{\Omega}\right)+\tilde{\Pi}\Omega\right\}.
\label{revision3}
\end{equation}
Also, the chemical density $x$ is a nonlinear function $\rho_\mathcal{X}(X)$ of $X$ as $x(X)=X/\Omega\left(X\right)=:\rho_\mathcal{X}(X)$.

In this case, the internal pressure $\varphi^{*}(y)$ is restricted to the constant external pressure $\tilde{\Pi}$, whereas the chemical equilibrium pressure $\varphi^{*}(y^\mathrm{EQ})$ is specified by the chemical potentials $\tilde{\mu}$ in the reservoir. 
If $\varphi^{*}(y)=\tilde{\Pi}$ does not balance with $\varphi^{*}(y^\mathrm{EQ})$, the system can not converge to the equilibrium state, and this imbalance drives growth or shrinking of the volume.
Whether growth or shrinking occurs is determined by the second law and the functional form of total entropy, which is represented for growing systems as 
\begin{equation}
    \Sigma^{\mathrm{t}\mathrm{o}\mathrm{t}}\left(X\right)=\frac{\Omega\left(X\right)}{\tilde{T}}K^{\mathcal{Y}}(y\left(X\right)) + \mathrm{const.},
\end{equation}
where $K^{\mathcal{Y}}(y)$ is defined as
\begin{equation}
    K^{\mathcal{Y}}(y) :=\varphi^{*}\left(y^{\mathrm{E}\mathrm{Q}}\right)-\tilde{\Pi}-\mathcal{D}^{\mathcal{Y}}\left[y^{\mathrm{E}\mathrm{Q}}||y \right].
\end{equation}
Here, $\mathcal{D}^{\mathcal{Y}}\left[y'||y \right]$ is the Bregman divergence \cite{c6,g1,g2} induced by $\varphi^{*}(y)$, and $y(X)=\rho^{\mathcal{Y}}\left(X\right)=\partial\varphi(\rho_\mathcal{X}(X))$ is a nonlinear map to associate the number of chemicals $X$ with a chemical potential $y$. 

\subsection{C. The conditions for growth, shrinking, and equilibration}
Our first claim provides the condition that determines the fate of the system, i.e., growth, shrinking or equilibration. 

\begin{remark}
%Consider a system equipped with an entropy density function $\sigma\left[\epsilon,n,x\right]$, and the regular stoichiometric matrix for the confined chemicals $S$; the reservoir is characterized by the intensive variables $(\tilde{T},\tilde{\Pi},\tilde{\mu})$. 
The fate of the system is classified by the sign of $\varphi^{*}( y^{\mathrm{E}\mathrm{Q}})-\tilde{\Pi}$ as follows:
\begin{enumerate}
    \item If and only if $\varphi^{*}\left( y^{\mathrm{E}\mathrm{Q}} \right)-\tilde{\Pi} = 0$, equilibrium states exist and the system converges to one of them. 
    \item If and only if $\varphi^{*}( y^{\mathrm{E}\mathrm{Q}} )-\tilde{\Pi} < 0$, 
the system eventually shrinks and finally vanishes. 
    \item If and only if $\varphi^{*}\left( y^{\mathrm{E}\mathrm{Q}} \right)-\tilde{\Pi} > 0$, the system is growing.
\end{enumerate}
\textbf{(see Sec. V and Theorem 1 for details)}.
\label{claim1}
\end{remark}
This result indicates that the system equilibrates only if the pressure $\tilde{\Pi}$ specified by the reservoir happens to coincide with the chemical equilibrium pressure $\varphi^{*}(y^{\mathrm{E}\mathrm{Q}})$ determined by the reservoir chemical potentials $\tilde{\mu}$.
Otherwise, the system shrinks or grows.

\textbf{ } 

\begin{figure}
    \centering
    \includegraphics[width=0.5\textwidth, clip]{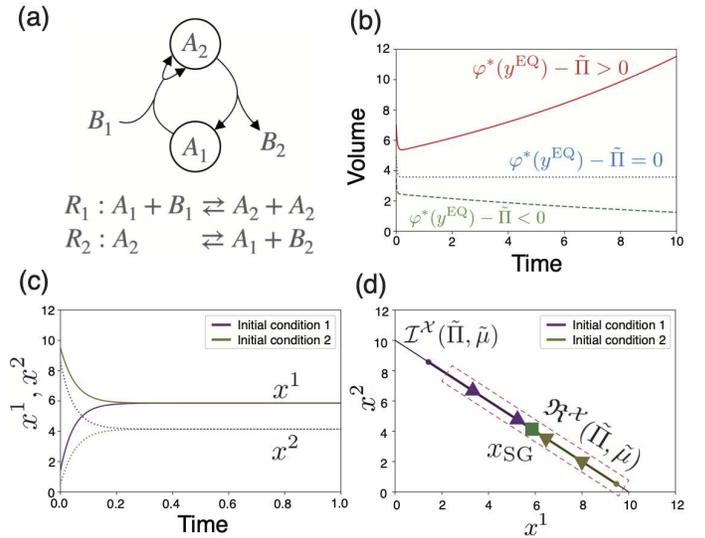}
    \caption{(a) A graph representation and chemical equations of a minimal motif of autocatalytic cycles. Two confined chemicals $A = (A_1,A_2)$ and two open chemicals $B=(B_1,B_2)$ undergo the two reactions $R_1$ and $R_2$. (b) The time evolution of the volume of the system for different parameter sets (see the caption in FIG. \ref{fig:numerical_simulation_sec8} for specific values of the parameters). The fate of the system is classified by the sign of $\varphi^*(y^{\mathrm{EQ}})-\tilde{\Pi}$. (c) The time evolution of the densities ($x^1$,$x^2$) of the confined chemicals ($A_1$,$A_2$) for the growth case $\varphi^*(y^{\mathrm{EQ}})-\tilde{\Pi}>0$. The evolutions are shown for two different initial conditions $1$ and $2$. (d) The trajectories of the system in the density space. They are constrained to the isobaric manifold $\mathcal{I}^{\mathcal{X}}(\tilde{\Pi},\tilde{\mu})$. In this example, the system converges to a steady growing state $x_{\mathrm{SG}}$ (green square), irrespective of the initial conditions. Such a steady growing state must be in the region $\mathfrak{R}^{\mathcal{X}}(\tilde{\Pi}, \tilde{\mu}) \subset \mathcal{I}^{\mathcal{X}}(\tilde{\Pi},\tilde{\mu})$, highlighted by the red dashed rectangle.}
    \label{fig:numerical_outline}
\end{figure}

\textit{Example 1:} To give an intuitive demonstration, we consider a minimal motif of autocatalytic cycles (see FIG. \ref{fig:numerical_outline}(a)). 
Here, two confined chemicals $A = (A_1, A_2)$ and two open chemicals $B = (B_1, B_2)$ are involved in the two reactions $R_1$ and $R_2$.  
We can regard the open chemicals $B_1$ and $B_2$ as a resource and a waste, respectively, because they are consumed and produced when the reactions forwardly progress.
The stoichiometric matrices can be represented as 
\begin{align}
    S = 
    \bordermatrix{     & R_1 & R_2 \cr
               A_1 & -1 & 1 \cr
               A_2 & 2 & -1 \cr
            },\mbox{  } 
    O = 
    \bordermatrix{     & R_1 & R_2 \cr
               B_1 & -1 & 0 \cr
               B_2 & 0 & 1
            }.
\end{align}
The regularity of the matrix $S$ is confirmed by $\det[S]=-1\neq0$.  
Denoting the number of $A = (A_1, A_2)$ by $X = (X^1, X^2)$, the reaction dynamics for the confined chemicals is written as
\begin{equation}
    \frac{dX^{i}}{dt}=S_{r}^{i}J^{r}\left(t\right).\label{SDynamics_example}
\end{equation}
In this example, we employ mass action kinetics with the local detailed balance condition \cite{sughiyama01,kobayashi01,07,06,02} for the reaction flux $J(t)$ \textbf{(see Sec. VIII for details)}.
Furthermore, we assume the ideal gas potential: the functional form of $\varphi^*(y)$ is obtained as 
\begin{eqnarray}
\displaystyle \varphi^{*}\left(y\right)&=&R\displaystyle \tilde{T}\left[\sum_{i}e^{\frac{y_{i}-\nu_{i}^{o}\left(\tilde{T}\right)}{R\tilde{T}}}+\sum_{m}e^{\frac{\tilde{\mu}_{m}-\mu_{m}^{o}\left(\tilde{T}\right)}{R\tilde{T}}}\right],
\label{revision2}
\end{eqnarray}
\textbf{(see Eq. (\ref{IGDTPg}) in Sec. VII for a derivation)}.
Then, by substituting the Legendre transformation of Eq. (\ref{revision2}) into Eq. (\ref{revision3}), we can calculate the volume $\Omega(X)$ as 
%The volume $\Omega(t)$ can be calculated by the equation of state, 
\begin{equation}
    \Omega(X) = \frac{R\tilde{T} \sum_i X^i}{\tilde{\Pi}-R\tilde{T} \sum_m \tilde{n}^m},\label{revision4}
\end{equation}
\textbf{(see Eq. (\ref{IG_ES}) in Sec. VII for details)} \cite{add_note1}.
This expression of the volume corresponds to the equation of state.
In FIG. \ref{fig:numerical_outline}(b), we verified Claim \ref{claim1} by numerical simulation. Indeed, the fate of the system is classified by the sign of $\varphi^*(y^{\mathrm{EQ}})-\tilde{\Pi}$. 

For the example of the ideal dilute solution, it is sufficient to just modify the standard chemical potentials $\nu^o(\tilde{T})$ and $\mu^o(\tilde{T})$ in Eq. (\ref{revision2}) \cite{07,thermo1}, because the solvent can be regarded as the background of the CRS. Then, Eq. (\ref{revision4}) can be read as Van Hoff’s law and $\tilde{\Pi}$ corresponds to the osmotic pressure. 
\rightline{$\square$}

%%%%%%%%%%%%%%%%%%%%%%%%%%%%%%%%%%%%%%%%%%%%%
\subsection{D. Thermodynamic constraint of isobaric dynamics}
Under isobaric conditions with a fast volume flux $J_\Omega(t)$, the pressure of the system should balance with the pressure of the reservoir. 
This constraint naturally defines the isobaric manifold in the chemical potential space:  
\begin{equation}
\mathcal{I}^{\mathcal{Y}}\left(\tilde{\Pi},\tilde{\mu}\right):=\left\{y|\varphi^{*}\left(y\right)-\tilde{\Pi} = 0\right\}.
\end{equation}
Its Legendre transform $\mathcal{I}^{\mathcal{X}}(\tilde{\Pi},\tilde{\mu}):=\partial \varphi^{*}(\mathcal{I}^{\mathcal{Y}})$ is a hypersurface in the density space.
Thus, $\mathcal{I}^{\mathcal{X}}(\tilde{\Pi},\tilde{\mu})$ and $\mathcal{I}^{\mathcal{Y}}(\tilde{\Pi},\tilde{\mu})$ characterize the thermodynamically admissible submanifolds in the density and chemical potential spaces, respectively.

\textbf{ } 

\textit{Example 2:} For the autocatalytic motif in FIG. \ref{fig:numerical_outline}(a), the time evolution of $x(t)$ is shown in FIG. \ref{fig:numerical_outline}(c) for the growth case in FIG. \ref{fig:numerical_outline}(b). 
This time evolution is actually constrained to the isobaric manifold  $\mathcal{I}^{\mathcal{X}}(\tilde{\Pi}, \tilde{\mu})$ as shown in FIG. \ref{fig:numerical_outline}(d). 
Since we have assumed ideal gas potentials, the isobaric manifold $\mathcal{I}^{\mathcal{X}}(\tilde{\Pi}, \tilde{\mu})$ reduces to a simplex by the equation of state \textbf{(see Sec. VII for details)}. 
\rightline{$\square$}

%%%%%%%%%%%%%%%%%%%%%%%%%%%%%%%%%%%%%%%%%%%%%%%%%%%%%%%%%%%%%
\subsection{E. The constraints and thermodynamic properties associated with the steady growing state}
Finally, we clarify the additional constraint imposed on the steady growing state $x_{\mathrm{SG}}$. 
The steady growing state is defined as a state such that the density remains constant with time whereas the volume keeps increasing \cite{muller02,muller01,Daan,Hidde}.
For the autocatalytic motif shown in FIG. \ref{fig:numerical_outline}(a), such a state $x_{\mathrm{SG}}$ exists and $x(t)$ converges to a steady growing state as in FIG. \ref{fig:numerical_outline}(c, d).

At this state, the entropy production rate can be expressed as
\begin{equation}
\dot{\Sigma}^{\mathrm{tot}}\left(\Omega\left(t\right)x_{\mathrm{S}\mathrm{G}}\right) = \frac{\dot{\Omega}(t)}{\tilde{T}} K^{\mathcal{Y}}\left(y^{\mathrm{SG}}\right), \label{entSGtot0}
\end{equation}
where $y^{\mathrm{SG}}$ is the Legendre transform of $x_{\mathrm{SG}}$ by $\partial\varphi$.
Because $\dot{\Omega}(t)>0$ at the growing state, $K^{\mathcal{Y}}(y^{\mathrm{SG}})$ should be positive by the second law. 
This requirement implies that $y^{\mathrm{SG}}$ should lie in the region 
$\mathfrak{R}^{\mathcal{Y}}(\tilde{\Pi},\tilde{\mu})=\mathcal{I}^{\mathcal{Y}}(\tilde{\Pi},\tilde{\mu})\cap\mathcal{Z}^\mathcal{Y}(\tilde{\mu})$. 
Here, 
\begin{equation}
    \mathcal{Z}^\mathcal{Y}\left(\tilde{\mu}\right):=\left\{y|\varphi^{*}\left(y^{\mathrm{E}\mathrm{Q}}\right)-\varphi^{*}\left(y\right)-\mathcal{D}^{\mathcal{Y}}\left[y^{\mathrm{E}\mathrm{Q}}||y \right]>0\right\},
\end{equation}
designates the region in which the positivity of entropy production rate is guaranteed.
By transferring this condition into the density space by the Legendre transformation, we have the following claim for $x_{\mathrm{SG}}$:
\begin{remark}
When $\varphi^{*}\left( y^{\mathrm{E}\mathrm{Q}} \right)-\tilde{\Pi} > 0$ and a steady growing state $x_{\mathrm{SG}}$ exists, 
the state $x_{\mathrm{SG}}$ must be in the region $\mathfrak{R}^{\mathcal{X}}(\tilde{\Pi},\tilde{\mu})$, where $\mathfrak{R}^{\mathcal{X}}(\tilde{\Pi},\tilde{\mu})=\partial \varphi^{*}(\mathfrak{R}^{\mathcal{Y}})$.
The entropy production rate at the state $x_{\mathrm{SG}}$ is represented as Eq. (\ref{entSGtot0}). 
\textbf{(See Sec. VI and Theorem 2 for the details)} 
%Then, the entropy production rate can be represented as 
%\begin{equation}
%    \dot{\Sigma}^{\mathrm{tot}}\left(t\right) = \frac{\dot{\Omega}(t)}{\tilde{T}} K^{\mathcal{X}}\left(x_{\mathrm{SG}}\right).
%\end{equation}
\label{claim2}
\end{remark}

\textbf{ } 

\textit{Example 3:} For the autocatalytic motif in FIG. \ref{fig:numerical_outline}(a), the steady growing state $x_{\mathrm{SG}}$ is indeed located within the region $\mathfrak{R}^{\mathcal{X}}(\tilde{\Pi},\tilde{\mu})$ (see FIG. \ref{fig:numerical_outline}(d)).

\begin{figure}
\includegraphics[width=0.5\textwidth, clip]{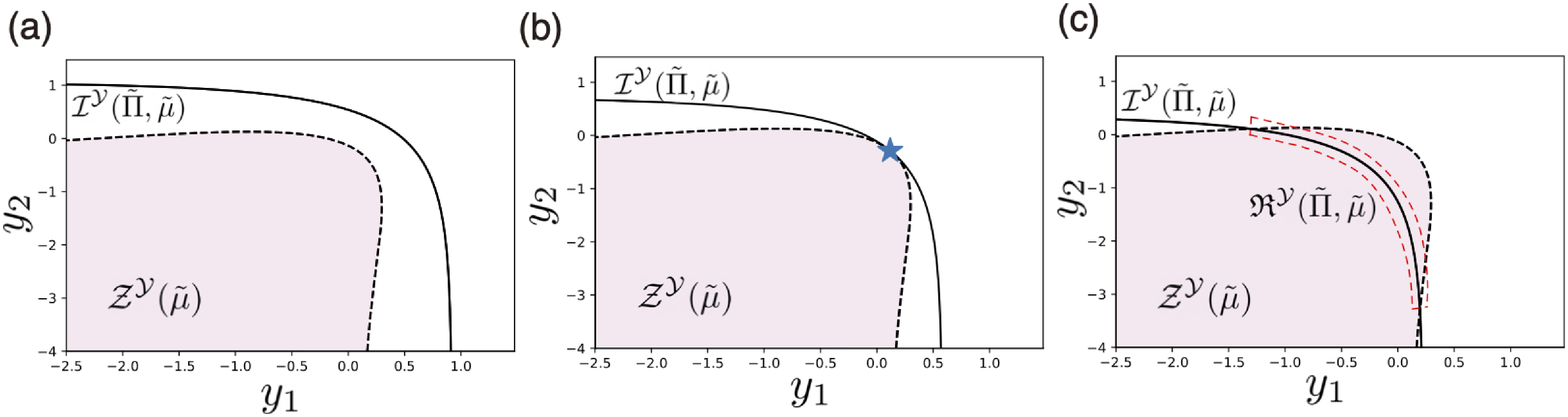}
\caption{Transition from the shrinking to the growing case. The isobaric manifold $\mathcal{I}^{\mathcal{Y}}(\tilde{\Pi},\tilde{\mu})$ and the region $\mathcal{Z}^\mathcal{Y}\left(\tilde{\mu}\right)$ in the chemical potential space are indicated by the solid curve and the light pink color, respectively. 
(a) When $\varphi^{*}( y^{\mathrm{E}\mathrm{Q}} )-\tilde{\Pi} < 0$ holds (i.e., the shrinking case), the intersection does not exist.
(b) When $\varphi^{*}( y^{\mathrm{E}\mathrm{Q}} )-\tilde{\Pi} = 0$ holds (i.e., the equilibrating case), transition from shrinking to growing occurs.
(c) When $\varphi^{*}( y^{\mathrm{E}\mathrm{Q}} )-\tilde{\Pi} > 0$ holds (i.e., the growing case), the intersection $\mathfrak{R}^{\mathcal{Y}}(\tilde{\Pi},\tilde{\mu})=\mathcal{I}^{\mathcal{Y}}(\tilde{\Pi},\tilde{\mu})\cap\mathcal{Z}^\mathcal{Y}(\tilde{\mu})$ exists, which is highlighted by the
curved red rectangle.}
\label{addfig}
\end{figure}

Moreover, we can verify that the transition from the shrinking to the growing case occurs when the intersection between $\mathcal{I}^{\mathcal{Y}}(\tilde{\Pi},\tilde{\mu})$ and $\mathcal{Z}^\mathcal{Y}\left(\tilde{\mu}\right)$ appears (see FIG. \ref{addfig}(b)).
In FIG. \ref{addfig}, the isobaric manifold $\mathcal{I}^{\mathcal{Y}}(\tilde{\Pi},\tilde{\mu})$ and the region $\mathcal{Z}^\mathcal{Y}\left(\tilde{\mu}\right)$ are indicated in the chemical potential space. 
For the shrinking case (FIG. \ref{addfig}(a)), the intersection is empty. 
By contrast, for the growing case (FIG. \ref{addfig}(c)), the intersection exists.  
\rightline{$\square$}

This concludes the outline of all our main results, which consist of the condition of growth, the constraints of growing systems and steady growing states, and the forms of total entropy and entropy production rate at the steady growing state.

\section{III. Thermodynamics for growing systems}
\label{SecIII}

From this section onward, we work on the derivation of our main claims introduced in Sec. II.
In this section, we derive the form of the total entropy, Eq. (\ref{enttot}), more specifically by employing time-scale separation. As a result, we will obtain the total entropy function for the reaction dynamics, Eq. (\ref{enttot_X}). 
Also, we will show that, given the number of the confined chemicals $X$, the volume $\Omega$ of the system is determined by the variational form, Eq. (\ref{OQEQ}), with the partial grand potential, Eq. (\ref{PGP}).   

Since we have assumed that $J_{E}\left(t\right),J_{\Omega}\left(t\right),J_{D}(t) \gg J\left(t\right)$,
we can analyze the dynamics, Eqs. (\ref{SDynamics}) and (\ref{RDynamics}), by separating the slow one $J(t)$ from the fast ones $J_{E}\left(t\right),J_{\Omega}\left(t\right),J_{D}(t)$. 
By solving the fast dynamics using the second law (see Appendix B), we obtain the effective slow dynamics (the reaction dynamics) as 
\begin{eqnarray}
\displaystyle \nonumber&&\frac{dX^{i}}{dt}=S_{r}^{i}J^{r}\left(t\right),\mbox{  }\frac{d\tilde{E}}{dt}=-\frac{dE_{\mathrm{Q}\mathrm{E}\mathrm{Q}}\left(X\right)}{dt},\\
\displaystyle \nonumber&&\frac{d\tilde{\Omega}}{dt}=-\frac{d\Omega_{\mathrm{Q}\mathrm{E}\mathrm{Q}}\left(X\right)}{dt},\mbox{  }\frac{d\tilde{N}^{m}}{dt}=O_{r}^{m}J^{r}\left(t\right)-\frac{dN_{\mathrm{Q}\mathrm{E}\mathrm{Q}}^{m}\left(X\right)}{dt},\\\label{ReactD}
\end{eqnarray}
where $\left(\cdot\right)_{\mathrm{Q}\mathrm{E}\mathrm{Q}}$ represents the value at the equilibrium state of the fast dynamics. We call this the quasi-equilibrium state, because it is not the equilibrium state of the slow dynamics. 
By using the partial grand potential:
\begin{equation}
\displaystyle \Phi\left[\tilde{T},\tilde{\mu};\Omega,X\right]:=\min_{E,N}\left\{E-\tilde{T}\Sigma\left[E,\Omega,N,X\right]-\tilde{\mu}_{m}N^{m}\right\},\label{PGP}
\end{equation}
the volume at the quasi-equilibrium state with the number of the confined chemicals $X$ can be evaluated by the variational form:
\begin{equation}
\displaystyle \Omega_{\mathrm{Q}\mathrm{E}\mathrm{Q}}\left(X\right)=\arg\min_{\Omega}\left\{\Phi\left[\tilde{T},\tilde{\mu};\Omega,X\right]+\tilde{\Pi}\Omega\right\}.\label{OQEQ}
\end{equation}
In addition, the other extensive variables can be calculated by differentiations of $\Phi[\tilde{T},\tilde{\mu};\Omega_{\mathrm{Q}\mathrm{E}\mathrm{Q}},X]$ as 
\begin{eqnarray}
\displaystyle \nonumber\Sigma_{\mathrm{Q}\mathrm{E}\mathrm{Q}}\left(X\right)&=&-\displaystyle \frac{\partial\Phi\left[\tilde{T},\tilde{\mu};\Omega_{\mathrm{Q}\mathrm{E}\mathrm{Q}},X\right]}{\partial\tilde{T}},\\
\displaystyle \nonumber N_{\mathrm{Q}\mathrm{E}\mathrm{Q}}^{m}\left(X\right)&=&-\displaystyle \frac{\partial\Phi\left[\tilde{T},\tilde{\mu};\Omega_{\mathrm{Q}\mathrm{E}\mathrm{Q}},X\right]}{\partial\tilde{\mu}_{m}},\\
\displaystyle \nonumber E_{\mathrm{Q}\mathrm{E}\mathrm{Q}}\left(X\right)&=&\displaystyle \Phi\left[\tilde{T},\tilde{\mu};\Omega_{\mathrm{Q}\mathrm{E}\mathrm{Q}},X\right]-\tilde{T}\frac{\partial\Phi\left[\tilde{T},\tilde{\mu};\Omega_{\mathrm{Q}\mathrm{E}\mathrm{Q}},X\right]}{\partial\tilde{T}}\\
&&-\displaystyle \tilde{\mu}_{m}\frac{\partial\Phi\left[\tilde{T},\tilde{\mu};\Omega_{\mathrm{Q}\mathrm{E}\mathrm{Q}},X\right]}{\partial\tilde{\mu}_{m}},\label{secIII:14}
\end{eqnarray}
where $\Sigma_{\mathrm{Q}\mathrm{E}\mathrm{Q}}\left(X\right)$ is the abbreviation of $\Sigma\left[E_{\mathrm{Q}\mathrm{E}\mathrm{Q}},\Omega_{\mathrm{Q}\mathrm{E}\mathrm{Q}},N_{\mathrm{Q}\mathrm{E}\mathrm{Q}},X\right]$.
The details of the derivation are shown in Appendix B. 
The formal solution of Eq. (\ref{ReactD}) with the initial condition $(X_{0},\tilde{E}\left(0\right),\tilde{\Omega}\left(0\right),\tilde{N}\left(0\right))$ is represented as 
\begin{eqnarray}
\nonumber X^{i}\left(t\right)&=&X_{0}^{i}+S_{r}^{i}\Xi^{r}\left(t\right),\\
\nonumber\tilde{E}\left(t\right)&=&\tilde{E}\left(0\right)-E_{\mathrm{Q}\mathrm{E}\mathrm{Q}}\left(X\left(t\right)\right),\\
\nonumber\tilde{\Omega}\left(t\right)&=&\tilde{\Omega}\left(0\right)-\Omega_{\mathrm{Q}\mathrm{E}\mathrm{Q}}\left(X\left(t\right)\right),\\
\tilde{N}^{m}\left(t\right)&=&\tilde{N}^{m}\left(0\right)+O_{r}^{m}\Xi^r\left(t\right)-N_{\mathrm{Q}\mathrm{E}\mathrm{Q}}^{m}\left(X\left(t\right)\right),\label{Sevo}
\end{eqnarray}
where $\Xi\left(t\right)=\left\{\Xi^{r}\left(t\right)\right\}$ is the integration of $J\left(t\right)$ with the initial condition $\Xi\left(0\right)=0$; this is known as the extent of reaction in chemistry. 
%, we note that the initial condition for the reservoir, $\left(\tilde{E}\left(0\right),\tilde{\Omega}\left(0\right),\tilde{N}\left(0\right)\right)$, coincides with the converged values of the fast dynamics (see Appendix A). 
Since we have assumed that $S$ is regular, 
there are no stoichiometric constraints that restrict attainable state of $X(t)$ by its initial state $X(0)$; i.e., the stoichiometric compatibility class \cite{m1, sughiyama01,kobayashi01} becomes $\mathbb{R}_{>0}^{\mathcal{N}_X}$.
Furthermore, by using the inverse matrix $S^{-1}$, the last equation in Eq. (\ref{Sevo}) can be rewritten as
\begin{equation}
\tilde{N}^{m}\left(t\right)=O_{r}^{m}\left(S^{-1}\right)_{i}^{r}X^{i}\left(t\right)-N_{\mathrm{Q}\mathrm{E}\mathrm{Q}}^{m}\left(X\left(t\right)\right)+\mathrm{const}.,\label{evotilN}
\end{equation}
where we substitute $\Xi^{r}\left(t\right)=\left(S^{-1}\right)_{i}^{r}\left\{X^{i}\left(t\right)-X_{0}^{i}\right\}$ into the last equation and abbreviate the terms composed of the initial condition to ``$\mathrm{const}.$". 
The representation of Eq. (\ref{evotilN}) implies that our reaction dynamics can be completely described only by the time evolution of the confined chemicals, $X\left(t\right)$. 

Next, we consider the time evolution of the total entropy during the reaction dynamics. 
By substituting Eqs. (\ref{Sevo}) and (\ref{evotilN}) into Eq. (\ref{enttot}), we obtain
\begin{eqnarray}
\displaystyle \nonumber&&\Sigma^{\mathrm{t}\mathrm{o}\mathrm{t}}\left(X\right)=\Sigma_{\mathrm{Q}\mathrm{E}\mathrm{Q}}\left(X\right)-\frac{1}{\tilde{T}}E_{\mathrm{Q}\mathrm{E}\mathrm{Q}}\left(X\right)-\frac{\tilde{\Pi}}{\tilde{T}}\Omega_{\mathrm{Q}\mathrm{E}\mathrm{Q}}\left(X\right)\\
\displaystyle \nonumber&&+\frac{\tilde{\mu}_{m}}{\tilde{T}}N_{\mathrm{Q}\mathrm{E}\mathrm{Q}}^{m}\left(X\right)-\frac{\tilde{\mu}_{m}}{\tilde{T}}O_{r}^{m}\left(S^{-1}\right)_{i}^{r}X^{i}+\mathrm{const}.\\
\displaystyle \nonumber&=&-\displaystyle \frac{1}{\tilde{T}}\left\{\Phi\left[\tilde{T},\tilde{\mu};\Omega_{\mathrm{Q}\mathrm{E}\mathrm{Q}},X\right]+\tilde{\Pi}\Omega_{\mathrm{Q}\mathrm{E}\mathrm{Q}}\left(X\right)-y_{i}^{\mathrm{E}\mathrm{Q}}X^{i}\right\}\\
&&+\mathrm{const}.,
\label{enttot_X}
\end{eqnarray}
where we employ the Taylor expansion for $\tilde{\Sigma}_{\tilde{T},\tilde{\Pi},\tilde{\mu}}$  and the partial grand potential, Eq. (\ref{PGP}); for simplicity, we also define
\begin{equation}
y_{i}^{\mathrm{E}\mathrm{Q}}:=-\tilde{\mu}_{m}O_{r}^{m}\left(S^{-1}\right)_{i}^{r}.
\label{y_EQ}
\end{equation}
The details of the derivation for Eqs. (\ref{enttot_X}) and (\ref{y_EQ}) are shown in Appendix B.

According to the second law, the system must climb up the landscape determined by the concave function $\Sigma^{\mathrm{t}\mathrm{o}\mathrm{t}}\left(X\right)$ \cite{n1} and finally converge to its maximum, which is called the equilibrium state, if it exists.
Therefore, to elucidate the fate of the system, it is important to analyze the form of the concave function $\Sigma^{\mathrm{t}\mathrm{o}\mathrm{t}}\left(X\right)$. 
We can briefly classify the form of $\Sigma^{\mathrm{t}\mathrm{o}\mathrm{t}}\left(X\right)$ into the following three cases: 
(1) If $\Sigma^{\mathrm{t}\mathrm{o}\mathrm{t}}\left(X\right)$ is bounded above and the points attaining its maximum are in the interior of the domain of $X$, i.e., $\arg\max_{X} \{\Sigma^{\mathrm{t}\mathrm{o}\mathrm{t}}\left(X\right)\} \in \mathbb{R}_{>0}^{\mathcal{N}_X}$, equilibrium states exist and the system converges to one of them.
(2) If $\Sigma^{\mathrm{t}\mathrm{o}\mathrm{t}}\left(X\right)$ is bounded above and the maximum of $\Sigma^{\mathrm{t}\mathrm{o}\mathrm{t}}\left(X\right)$ is at $X = 0$, the volume $\Omega_{\mathrm{Q}\mathrm{E}\mathrm{Q}}\left(X(t)\right)$ eventually shrinks and finally vanishes. 
(3) If $\Sigma^{\mathrm{t}\mathrm{o}\mathrm{t}}\left(X\right)$ is not bounded above, $X(t)$ diverges in the reaction dynamics.
Also, the volume $\Omega_{\mathrm{Q}\mathrm{E}\mathrm{Q}}\left(X\right)$ diverges for $X\rightarrow\infty$, because of the homogeneity of the volume. 
This situation corresponds to the growth of the system. 
The main aim of this work is to reveal what condition distinguishes these three cases. 
In the remaining part of this paper, we will address this problem by employing Hessian and projective geometry. 

\section{IV. Preparation for a geometric representation of isobaric chemical reaction systems}
\label{SecIV}
\begin{figure}
    \centering
    \includegraphics[width=0.5\textwidth, clip]{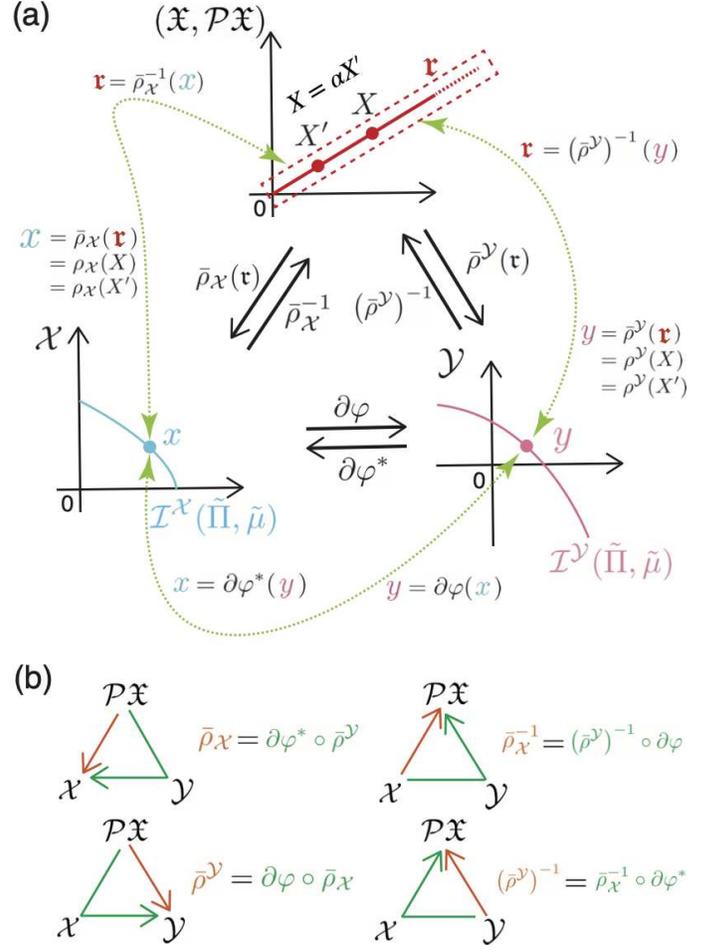}
    \caption{Diagrammatic representation of the triad of spaces, ($\mathfrak{X},\mathcal{P}\mathfrak{X}$), $\mathcal{X}$, and $\mathcal{Y}$. (a) The top space, $\mathfrak{X} = \mathbb{R}_{>0}^{\mathcal{N}_X}$, represents the number of the confined chemicals $X$. We also define the set of rays in $\mathfrak{X}$ as $\mathcal{P}\mathfrak{X}$. 
    An element $\mathfrak{r} \in \mathcal{P}\mathfrak{X}$ is a ray, which is a subset of $\mathfrak{X}$.
    %a subset of $\mathfrak{X}$ such that any two points $X$ and $X'$ on the ray $\mathfrak{r}$ satisfy $X = \alpha X'$ $(\alpha > 0)$. 
    The spaces on the left and right bottom represent the density space $\mathcal{X}= \mathbb{R}_{>0}^{\mathcal{N}_X}$ and the chemical potential space $\mathcal{Y}= \mathbb{R}^{\mathcal{N}_X}$, respectively. 
    A ray $\mathfrak{r} \in \mathcal{P}\mathfrak{X}$ and a point $x$ in the isobaric manifold $\mathcal{I}^{\mathcal{X}}(\tilde{\Pi}, \tilde{\mu}) \subset \mathcal{X}$ are mapped to each other by $\bar{\rho}_{\mathcal{X}}(\mathfrak{r})$ and $\bar{\rho}^{-1}_{\mathcal{X}}(x)$. 
    Similarly, a ray $\mathfrak{r}$ and a point $y$ in the isobaric manifold $\mathcal{I}^{\mathcal{Y}}(\tilde{\Pi}, \tilde{\mu}) \subset \mathcal{Y}$ are mapped to each other by $\bar{\rho}^{\mathcal{Y}}(\mathfrak{r})$ and $\left(\bar{\rho}^{\mathcal{Y}}\right)^{-1}(y)$.
    The spaces, $\mathcal{X}$ and $\mathcal{Y}$, are mapped to each other by $\partial \varphi$ and $\partial \varphi^*$. 
    (b) The map $\bar{\rho}_{\mathcal{X}}$ from $\mathcal{P}\mathfrak{X}$ to $\mathcal{X}$ and its inverse $\bar{\rho}^{-1}_{\mathcal{X}}$ can be represented by the composition of two maps via the space $\mathcal{Y}$ (top line). Similarly, the map  $\bar{\rho}^{\mathcal{Y}}$ from $\mathcal{P}\mathfrak{X}$ to $\mathcal{Y}$ and its inverse $\left(\bar{\rho}^{\mathcal{Y}}\right)^{-1}$ are expressed by the composition of two maps via the space $\mathcal{X}$ (bottom line).  }
    \label{fig:trinity}
\end{figure}

We devote this section to preparation for the geometric representation of our system. 
As a result, it is revealed that any thermodynamic state is constrained to the isobaric manifolds $\mathcal{I}^\mathcal{X}(\tilde{\Pi},\tilde{\mu})$ and $\mathcal{I}^\mathcal{Y}(\tilde{\Pi},\tilde{\mu})$ in the density space $\mathcal{X}$ and the chemical potential space $\mathcal{Y}$, respectively.
Furthermore, we find a one-to-one correspondence between a density $x\in\mathcal{I}^\mathcal{X}(\tilde{\Pi},\tilde{\mu})$, a chemical potential $y\in\mathcal{I}^\mathcal{Y}(\tilde{\Pi},\tilde{\mu})$ and a ray $\mathfrak{r}$ in the number space $\mathfrak{X}$, as illustrated in Fig. \ref{fig:trinity}.

As mentioned in Sec. II, the homogeneity of the system entropy function allows us to write it as 
\begin{equation}
\Sigma\left[E,\Omega,N,X\right]=\Omega\sigma\left[\epsilon,n,x\right],
\end{equation}
where $\sigma\left[\epsilon,n,x\right]$ represents the entropy density and $\left(\epsilon,n, x\right):=\left(E/\Omega,N/\Omega,X/\Omega\right)$; also, we have assumed that $\sigma\left[\epsilon,n,x\right]$ is strictly concave. 
We introduce the number and the density spaces of the confined chemicals as $X\in \mathfrak{X}=\mathbb{R}_{>0}^{\mathcal{N}_X}$ and $x\in \mathcal{X}=\mathbb{R}_{>0}^{\mathcal{N}_X}$, respectively. 
Also, we define the partial grand potential density as $\varphi\left(x\right)=\varphi[\tilde{T},\tilde{\mu};x]:=\Omega^{-1}\Phi[\tilde{T},\tilde{\mu};\Omega,X]=\Phi[\tilde{T},\tilde{\mu};1,X/\Omega]$, where we use the homogeneity of $\Phi$. 
From the definition of $\Phi$, Eq. (\ref{PGP}), $\varphi[\tilde{T},\tilde{\mu};x]$ can be represented by a variant of the Legendre transformation of $\sigma\left[\epsilon,n,x\right]$ as
\begin{equation}
\displaystyle \varphi\left[\tilde{T},\tilde{\mu};x\right]=\min_{\epsilon,n}\left\{\epsilon-\tilde{T}\sigma\left[\epsilon,n,x\right]-\tilde{\mu}_{m}n^{m}\right\},
\label{PGPx}
\end{equation}
and therefore $\varphi\left(x\right)$ is strictly convex. 
By using $\varphi\left(x\right)$, we can rewrite Eq. (\ref{OQEQ}) as 
\begin{equation}
\displaystyle \Omega\left(X\right)=\Omega_{\mathrm{Q}\mathrm{E}\mathrm{Q}}\left(X\right)=\arg\min_{\Omega}\left\{\Omega\varphi\left(\frac{X}{\Omega}\right)+\tilde{\Pi}\Omega\right\}.\label{Ophi}
\end{equation}
For notational simplicity, we omit the subscript $\left(\cdot\right)_{\mathrm{Q}\mathrm{E}\mathrm{Q}}$, hereafter. 
Due to the strict convexity of $\varphi\left(x\right)$, the volume $\Omega\left(X\right)$ uniquely exists for any given $X$ (see Appendix C). 

The equation (\ref{Ophi}) implies that any possible state in the density space $\mathcal{X}$ is constrained to a submanifold as follows. 
The critical equation of Eq. (\ref{Ophi}) is given by
\begin{equation}
\displaystyle \varphi\left(\frac{X}{\Omega}\right)-\frac{X^{i}}{\Omega}\partial_{i}\varphi\left(\frac{X}{\Omega}\right)+\tilde{\Pi}=0,
\end{equation}
where $\partial_{i}\varphi\left(X/\Omega\right)=\left.\partial\varphi\left(x\right)/\partial x^{i}\right|_{x=X/\Omega}$. 
Therefore, any possible state lies in an isobaric manifold: 
\begin{equation}
\mathcal{I}^{\mathcal{X}}\left(\tilde{\Pi},\tilde{\mu}\right):=\left\{x|\varphi\left(x\right)-x^{i}\partial_{i}\varphi\left(x\right)+\tilde{\Pi}=0\right\}\subset\mathcal{X}.
\label{ibm_X}
\end{equation}
In other words, the time evolution of the density $x\left(t\right)$, given by Eq. (\ref{ReactD}), is constrained to this submanifold (see the left bottom panel in FIG. \ref{fig:trinity}(a)).

Next, we relate the number $X$ with the density $x$. To do this, we define a map from the number space $\mathfrak{X}$ to the isobaric manifold $\mathcal{I}^{\mathcal{X}}(\tilde{\Pi},\tilde{\mu})$: 
\begin{equation}
\displaystyle \rho_\mathcal{X}: X\in \mathfrak{X}\mapsto\rho_\mathcal{X}\left(X\right)=\left\{x^{i}\left(X\right)\right\}=\left\{\frac{X^{i}}{\Omega\left(X\right)}\right\} \in \mathcal{I}^{\mathcal{X}}.
\label{map_rho_X}
\end{equation}
This map gives the density of the confined chemicals at a quasi-equilibrium state with $X$.
Note that the map $\rho_\mathcal{X}$ is not injective because of the homogeneity: $\Omega\left(\alpha X\right)=\alpha\Omega\left(X\right)$ ($\alpha>0$), which is guaranteed by Eq. (\ref{Ophi}). 
This means that the map $\rho_\mathcal{X}$ satisfies 
\begin{equation}
\rho_\mathcal{X}\left(\alpha X\right)=\rho_\mathcal{X}\left(X\right), 
\label{rho_homo}
\end{equation}
and thus any point $X$ on a ray $\mathfrak{r}$ in the number space $\mathfrak{X}$ gives the same density $x=\rho_\mathcal{X}(X)$ (see the top panel in FIG. \ref{fig:trinity}(a)). 
The mathematical expression of the ray is given in \cite{n6}.

Here, the geometric representation of the map $\rho_\mathcal{X}$ is clarified by regarding the number space $\mathfrak{X}$ as a collection of rays; that is, we denote the projective space of $\mathfrak{X}$ as $ \mathcal{P}\mathfrak{X}$.
Elements of the space $\mathcal{P}\mathfrak{X}$ are rays $\mathfrak{r} \subset \mathfrak{X}$. 
Due to Eq. (\ref{rho_homo}), the map $\rho_\mathcal{X}$ descends to a well-defined map from $\mathcal{P}\mathfrak{X}$ to $\mathcal{I}^{\mathcal{X}}(\tilde{\Pi},\tilde{\mu})$:
\begin{equation}
\bar{\rho}_\mathcal{X}: \mathfrak{r}\in \mathcal{P}\mathfrak{X}\mapsto\bar{\rho}_\mathcal{X}\left(\mathfrak{r}\right)=\rho_\mathcal{X}\left(X\right) \in \mathcal{I}^{\mathcal{X}},\mbox{ }\mathrm{for}\mbox{ }X\in \mathfrak{r}.
\end{equation}
This map $\bar{\rho}_\mathcal{X}$ become injective \cite{n7}.
For a later analysis, we also define the inverse map of $\bar{\rho}_\mathcal{X}$ as $\bar{\rho}_\mathcal{X}^{-1}: \mathcal{I}^{\mathcal{X}}(\tilde{\Pi},\tilde{\mu})\rightarrow\mathcal{P}\mathfrak{X}$, which gives the corresponding ray to a given density $x\in \mathcal{I}^{\mathcal{X}}(\tilde{\Pi},\tilde{\mu})$ (see FIG. \ref{fig:trinity}(a)). 

Finally, we introduce the dual space of the density space $\mathcal{X}$ as $y\in \mathcal{Y}=\mathbb{R}^{\mathcal{N}_X}$. It is thermodynamically interpreted as the space of chemical potentials. 
Also, we define a map from $\mathcal{X}$ to $\mathcal{Y}$ by using the convex function $\varphi\left(x\right)$ as 
\begin{equation}
\partial\varphi: x\in \mathcal{X}\mapsto\partial\varphi\left(x\right)=\left\{\partial_{i}\varphi\right\}=\left\{\frac{\partial\varphi}{\partial x^{i}}\right\}\in \mathcal{Y},\label{mapXtoY}
\end{equation}
which outputs the value of chemical potential at a state $x$. 
Since $\varphi\left(x\right)$ is strictly convex, the map $\partial\varphi$ is injective. 
%Furthermore, for ordinary chemical reaction systems,  $\mathrm{Ran}[\partial\varphi]$ is $\mathbb{R}^{\mathcal{N}_X}$; thus $\partial\varphi$ is bijective. 
To construct the inverse map of $\partial\varphi$, we define the strictly convex function $\varphi^{*}\left(y\right)$ on the dual space $\mathcal{Y}$ by the Legendre transformation: 
\begin{equation}
\displaystyle \varphi^{*}\left(y\right):=\max_{x}\left\{y_{i}x^{i}-\varphi\left(x\right)\right\},\label{LTphi}
\end{equation}
which corresponds to the full grand potential density and gives a pressure of the system at a state $y$. 
Employing $\varphi^{*}\left(y\right)$, we can represent the inverse map as 
\begin{equation}
\partial\varphi^{*}: y\in \mathcal{Y}\mapsto\partial\varphi^{*}\left(y\right)=\left\{\partial^{i}\varphi^{*}\right\}=\left\{\frac{\partial\varphi^{*}}{\partial y_{i}}\right\}\in \mathcal{X}.\label{mapYtoX}
\end{equation}
These two spaces, $\mathcal{X}$ and $\mathcal{Y}$, together with the pair of convex functions, $\varphi(x)$ and $\varphi^*(y)$, constitute the Hessian geometric structure of chemical thermodynamics \cite{sughiyama01}. 
The structure is fundamental to capture a geometric relation between the two dual spaces and will be used intensively in the following sections.

The isobaric manifold $\mathcal{I}^{\mathcal{X}}(\tilde{\Pi},\tilde{\mu})$ in $\mathcal{X}$ is mapped via $\partial \varphi$ to the chemical potential space $\mathcal{Y}$ as 
\begin{equation}
\mathcal{I}^{\mathcal{Y}}\left(\tilde{\Pi},\tilde{\mu}\right):=\partial\varphi\left(\mathcal{I}^{\mathcal{X}}\right)=\left\{y|\varphi^{*}\left(y\right)-\tilde{\Pi} = 0\right\},
\label{ibm_Y}
\end{equation}
which is a level hypersurface for the dual convex function $\varphi^{*}(y)$ (see the right bottom panel in FIG. \ref{fig:trinity}(a)).
In addition, we define the map from $\mathfrak{X}$ to $\mathcal{I}^{\mathcal{Y}}(\tilde{\Pi},\tilde{\mu}) \subset \mathcal{Y}$ as $\rho^{\mathcal{Y}}(X):= \partial \varphi \circ \rho_{\mathcal{X}}(X)=y(X)$, which also induces the map:
\begin{equation}
\bar{\rho}^{\mathcal{Y}}: \mathfrak{r}\in \mathcal{P}\mathfrak{X}\mapsto\bar{\rho}^\mathcal{Y}\left(\mathfrak{r}\right)=\rho^\mathcal{Y}\left(X\right)\in \mathcal{I}^{\mathcal{Y}},\mbox{ }\mathrm{for}\mbox{ }X\in \mathfrak{r}.    
\end{equation}
Since this map is injective, we define inverse map as $\left(\bar{\rho}^{\mathcal{Y}}\right)^{-1}(y) = \bar{\rho}_{\mathcal{X}}^{-1} \circ \partial \varphi^* (y)$ (see FIG. \ref{fig:trinity}(b)).
The fact that the isobaric manifold is identical to a level hypersurface for a potential function is one of the fundamental constituents in the Hessian geometry. 

\section{V. Form of the total entropy function and the fate of the system}
\label{SecV}

With the preparation in the previous section, we are in the position to reveal 
the form of the total entropy function, Eq. (\ref{enttot_X}), and predict the fate of the system. 

For this purpose, we introduce the Bregman divergence \cite{c6,g1,g2} on $\mathcal{Y}$: 
\begin{equation}
\mathcal{D}^{\mathcal{Y}}\left[y||y^{\prime}\right]:=\left\{\varphi^{*}\left(y\right)-\varphi^{*}\left(y^{\prime}\right)\right\}-\partial^{i}\varphi^{*}\left(y^{\prime}\right)\left\{y_{i}-y_{i}^{\prime}\right\},\label{BDY}
\end{equation}
which measures the deviation at the point $y$ between the convex function $\varphi^*\left(y\right)$ and the tangent plane at the point $y^{\prime}$.
This divergence has the following property: $\mathcal{D}^{\mathcal{Y}}\left[y||y^{\prime}\right]\geq 0$, the equality holds if and only if $y=y^{\prime}$ and therefore it acts as an asymmetric distance from $y^{\prime}$ to $y$. 
The Bregman divergence is also one of the fundamental constitutes of Hessian geometry.

We rewrite the total entropy function Eq. (\ref{enttot_X}) by using the divergence as follows.  
Using the partial grand potential density $\varphi(x)$, Eq. (\ref{enttot_X}) can be rewritten as 
\begin{equation}
   \Sigma^{\mathrm{t}\mathrm{o}\mathrm{t}}\left(X\right)=-\displaystyle \frac{\Omega\left(X\right)}{\tilde{T}}\left\{\varphi\left(x\left(X\right)\right)-y_{i}^{\mathrm{E}\mathrm{Q}}x^{i}\left(X\right)+\tilde{\Pi}\right\},\label{ent41}
\end{equation}
where $x(X)=\rho_{\mathcal{X}}(X)$ is defined in Eq. (\ref{map_rho_X}) and we neglect the constant term. 
This equation is further rearranged as 
\begin{eqnarray}\nonumber
    \Sigma^{\mathrm{t}\mathrm{o}\mathrm{t}}\left(X\right)&=&\displaystyle \frac{\Omega\left(X\right)}{\tilde{T}}\left\{y_{i}^{\mathrm{E}\mathrm{Q}}-\partial_{i}\varphi\left(x\left(X\right)\right)\right\}x^{i}\left(X\right) \\
    &=& \frac{\Omega\left(X\right)}{\tilde{T}}\left\{y_{i}^{\mathrm{E}\mathrm{Q}}-y_i\left(X\right)\right\}\partial^{i}\varphi^{*}\left(y\left(X\right)\right).\label{ent27}
\end{eqnarray}
To derive the first line, we used Eq. (\ref{ibm_X}); in the second line, we employed the fact that the density $x$ and the chemical potential $y$ are mapped to each other by $\partial\varphi$ and $\partial\varphi^*$ (see FIG. \ref{fig:trinity}). 
Finally, using the Bregman divergence from $y(X)$ to $y^{\mathrm{E}\mathrm{Q}}$, we obtain 
\begin{equation}
\Sigma^{\mathrm{t}\mathrm{o}\mathrm{t}}\left(X\right)=\frac{\Omega\left(X\right)}{\tilde{T}}\left\{\varphi^{*}\left(y^{\mathrm{E}\mathrm{Q}}\right)-\tilde{\Pi}-\mathcal{D}^{\mathcal{Y}}\left[y^{\mathrm{E}\mathrm{Q}}||y\left(X\right)\right]\right\},\label{ent_tot_bd}
\end{equation}
where we employ $\varphi^*(y(X)) = \tilde{\Pi}$, because $y(X)=\rho^{\mathcal{Y}}(X) \in \mathcal{I}^{\mathcal{Y}}(\tilde{\Pi},\tilde{\mu})$ (see Eq. (\ref{ibm_Y})).
Here, we note that the first two terms in Eq. (\ref{ent_tot_bd}), $\varphi^{*}\left(y^{\mathrm{E}\mathrm{Q}}\right)-\tilde{\Pi}$, are calculated by the intensive variables of the reservoir, 
because $y^{\mathrm{E}\mathrm{Q}}$ is given by its chemical potential $\tilde{\mu}$ as in Eq. (\ref{y_EQ}). In the following, we will show that the sign of $\varphi^{*}\left(y^{\mathrm{E}\mathrm{Q}}\right)-\tilde{\Pi}$ determines the fate of the system. 

For convenience, we also denote terms in the bracket in Eq. (\ref{ent_tot_bd}) by 
\begin{equation}
    K^{\mathcal{Y}}(y) :=\varphi^{*}\left(y^{\mathrm{E}\mathrm{Q}}\right)-\tilde{\Pi}-\mathcal{D}^{\mathcal{Y}}\left[y^{\mathrm{E}\mathrm{Q}}||y \right],
    \label{Kappa_Y}
\end{equation}
that is, $\Sigma^{\mathrm{t}\mathrm{o}\mathrm{t}}\left(X\right)=\{\Omega\left(X\right)/\tilde{T}\}K^{\mathcal{Y}}(y(X))$.
Here, we emphasize that the value $K^{\mathcal{Y}}(y(X))$ is kept constant on each ray $\mathfrak{r}$ in the number space $\mathfrak{X}$, because $y(X)=\rho^{\mathcal{Y}}(X)=\mathrm{const.}$ for $X \in \mathfrak{r} \in \mathcal{P}\mathfrak{X}$.

First, let us consider the case $\varphi^{*}\left(y^{\mathrm{E}\mathrm{Q}}\right)-\tilde{\Pi}=0$, which corresponds to the situation that equilibrium states exist and the system converges to one of them. 
In this case, since $K^{\mathcal{Y}}(y) =-\mathcal{D}^{\mathcal{Y}}\left[y^{\mathrm{E}\mathrm{Q}}||y \right]$ and $\Omega(X) > 0$, the entropy function in Eq. (\ref{ent_tot_bd}) satisfies  $\Sigma^{\mathrm{t}\mathrm{o}\mathrm{t}}\left(X\right) \leq 0$, the equality holds if and only if  $y = y^{\mathrm{E}\mathrm{Q}}$.
Furthermore, from Eq. (\ref{ibm_Y}), $y^{\mathrm{E}\mathrm{Q}} \in \mathcal{I}^{\mathcal{Y}}(\tilde{\Pi},\tilde{\mu})$, and therefore $y(X)=\rho^{\mathcal{Y}}\left(X\right)$ can reach $y^{\mathrm{E}\mathrm{Q}}$.
Hence, the maximum of the entropy function is achieved on the ray given by $\left(\bar{\rho}^{\mathcal{Y}}\right)^{-1}(y^{\mathrm{E}\mathrm{Q}})$, which represents a set of the equilibrium states. 
Since the second law imposes that the total entropy function increases in the time evolution of the system, it will converge to a point on the equilibrium ray, depending on the initial condition and the functional form of the reaction flux $J(t)$ in Eq. (\ref{ReactD}).  
We should note that the equilibrium state is identified by a unique point in the density space $\mathcal{X}$. 
However, in the number space $\mathfrak{X}$, the equilibrium states form a ray and the equilibrium point to which the system converges is one of the points on the ray. 

\textbf{ } 

\begin{figure}
\includegraphics[width=0.5\textwidth, clip]{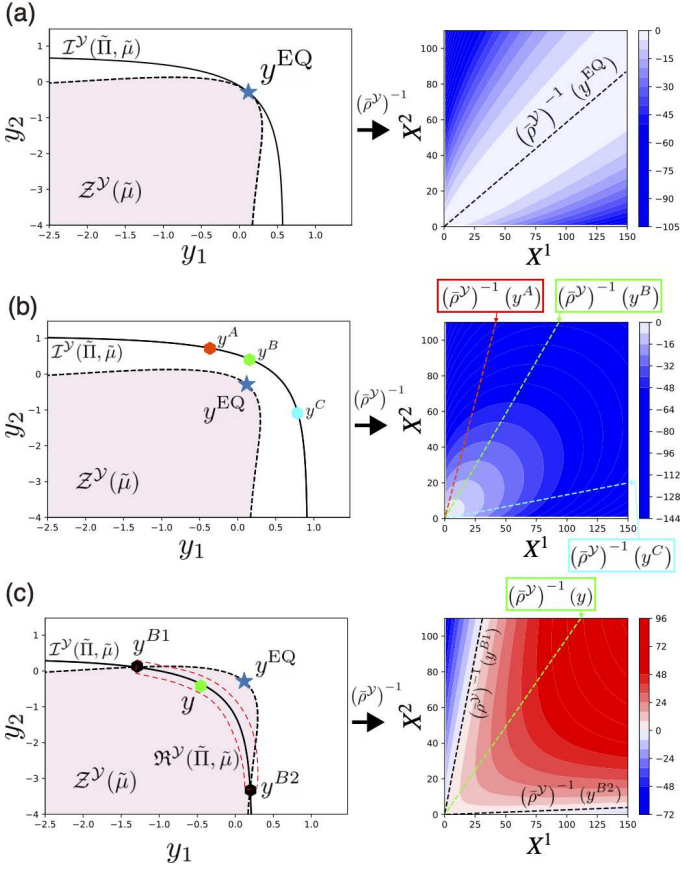}
\caption{For the autocatalytic motif shown in FIG. \ref{fig:numerical_outline}(a), we describe the isobaric manifold $\mathcal{I}^{\mathcal{Y}}(\tilde{\Pi},\tilde{\mu})$ in  $\mathcal{Y}$ (left panels) and the corresponding rays (right panels) in $\mathfrak{X}$ given by the map $\left(\bar{\rho}^{\mathcal{Y}}\right)^{-1} = \bar{\rho}_{\mathcal{X}}^{-1} \circ \partial \varphi^*$. The heat maps in the right panels indicate values of the entropy function $\Sigma^{\mathrm{t}\mathrm{o}\mathrm{t}}$ (see the caption in FIG. \ref{fig:numerical_simulation_sec8} for specific values of the parameters).  
(a) When $\varphi^*(y^{\mathrm{E}\mathrm{Q}}) - \tilde{\Pi} = 0$, the point $y^{\mathrm{E}\mathrm{Q}}$ lies in $\mathcal{I}^{\mathcal{Y}}(\tilde{\Pi}, \tilde{\mu})$ and $K^{\mathcal{Y}}(y^{\mathrm{E}\mathrm{Q}}) = 0$; for the other $y \in \mathcal{I}^{\mathcal{Y}}(\tilde{\Pi}, \tilde{\mu})$, the value of $K^{\mathcal{Y}}(y)$ is negative.
Thus, the maximum $\Sigma^{\mathrm{t}\mathrm{o}\mathrm{t}} = 0$ is achieved on the ray given by $\left(\bar{\rho}^{\mathcal{Y}}\right)^{-1}(y^{\mathrm{E}\mathrm{Q}})$. 
(b) When $\varphi^*(y^{\mathrm{E}\mathrm{Q}}) - \tilde{\Pi} < 0$, the point $y^{\mathrm{E}\mathrm{Q}}$ does not exist on $\mathcal{I}^{\mathcal{Y}}(\tilde{\Pi}, \tilde{\mu})$ and $K^{\mathcal{Y}}(y)$ is negative for all $y\in \mathcal{I}^{\mathcal{Y}}(\tilde{\Pi}, \tilde{\mu})$. Then, on a ray in $\mathfrak{X}$ given by $\left(\bar{\rho}^{\mathcal{Y}}\right)^{-1}(y)$, the value of $K^{\mathcal{Y}}(y)$ is negative and constant. 
Thus, on each ray, the entropy function $\Sigma^{\mathrm{t}\mathrm{o}\mathrm{t}}$ increases when $X$ approaches the origin. 
As a guide, we display typical points $y^A$, $y^B$ and $y^C$, and the corresponding rays in $\mathfrak{X}$.
(c) The region $\mathcal{Z}^\mathcal{Y}(\tilde{\mu})$ is indicated by light pink color in the left panel. 
Only when $\varphi^* (y^{\mathrm{EQ}})-\tilde{\Pi}>0$, the intersection  $\mathfrak{R}^{\mathcal{Y}}(\tilde{\Pi},\tilde{\mu})=\mathcal{I}^{\mathcal{Y}}(\tilde{\Pi},\tilde{\mu})\cap\mathcal{Z}^\mathcal{Y}(\tilde{\mu})$ appears, where $K^{\mathcal{Y}}(y)$ is positive for any $y \in \mathfrak{R}^{\mathcal{Y}}(\tilde{\Pi},\tilde{\mu})$. 
We highlight $\mathfrak{R}^{\mathcal{Y}}(\tilde{\Pi},\tilde{\mu})$ by the curved red rectangle, i.e., within the range between $y^{B1}$ and $y^{B2}$ in $\mathcal{I}^{\mathcal{Y}}(\tilde{\Pi},\tilde{\mu})$. 
Thus, on a ray $\left(\bar{\rho}^{\mathcal{Y}}\right)^{-1}(y)$ in $\mathfrak{X}$ for every $y \in \mathfrak{R}^{\mathcal{Y}}(\tilde{\Pi},\tilde{\mu})$, the entropy function $\Sigma^{\mathrm{t}\mathrm{o}\mathrm{t}}$ increases when $X$ diverges along the ray.  
We also show the points $y^{B1}$ and $y^{B2}$ at which $K^{\mathcal{Y}}(y)=0$, and the corresponding rays on which $\Sigma^{\mathrm{t}\mathrm{o}\mathrm{t}}=0$.} 
\label{fig:numerical_simulation_sec5}
\end{figure}

\textit{Example 4:} Consider the autocatalytic motif shown in FIG. \ref{fig:numerical_outline}(a) and the intensive variables $\tilde{\Pi}$ and $\tilde{\mu}$ in the reservoir satisfy $\varphi^* (y^{\mathrm{EQ}})-\tilde{\Pi}=0$.
In this case, the isobaric manifold $\mathcal{I}^{\mathcal{Y}}(\tilde{\Pi},\tilde{\mu})$ in the chemical potential space $\mathcal{Y}$ is shown in FIG. \ref{fig:numerical_simulation_sec5}(a), and $y^{\mathrm{EQ}}$ lies on $\mathcal{I}^{\mathcal{Y}}(\tilde{\Pi},\tilde{\mu})$. 
Furthermore, the maximum of the entropy function $\Sigma^{\mathrm{tot}}(X)$ is achieved on the ray given by $\left(\bar{\rho}^{\mathcal{Y}}\right)^{-1}(y^{\mathrm{E}\mathrm{Q}})$ (see the right panel of FIG. \ref{fig:numerical_simulation_sec5}(a)).

\rightline{$\square$}

Second, we show that the system eventually shrinks if $\varphi^{*}\left(y^{\mathrm{E}\mathrm{Q}}\right)-\tilde{\Pi}<0$. 
In this case, $K^{\mathcal{Y}}(y)$ is negative for all $y \in \mathcal{Y}$.
Thus, on a ray in $\mathfrak{X}$ given by $\left(\bar{\rho}^{\mathcal{Y}}\right)^{-1}(y)$ for any $y$, 
the value $K^{\mathcal{Y}}(y(X))$ is a negative constant. 
In addition, $\Omega(X)$ is an increasing function on the ray because of its homogeneity. 
Thus, the entropy function $\Sigma^{\mathrm{t}\mathrm{o}\mathrm{t}}$ increases when $X$ approaches the origin along the ray. Accordingly, the maximum of the entropy function (to be more precise, the supremum of the entropy function) is located at $X=0$; that is, the system eventually shrinks and finally vanishes. 

\textbf{ } 

\textit{Example 5:} For the autocatalytic motif shown in FIG. \ref{fig:numerical_outline}(a) under the condition $\varphi^* (y^{\mathrm{EQ}})-\tilde{\Pi}<0$, the isobaric manifold $\mathcal{I}^{\mathcal{Y}}(\tilde{\Pi},\tilde{\mu})$ in $\mathcal{Y}$ is shown in FIG. \ref{fig:numerical_simulation_sec5}(b). In this case, $y^{\mathrm{EQ}}$ does not exist on $\mathcal{I}^{\mathcal{Y}}(\tilde{\Pi},\tilde{\mu})$. For every $y \in \mathcal{I}^{\mathcal{Y}}(\tilde{\Pi},\tilde{\mu})$, the corresponding ray in $\mathfrak{X}$ is given by 
$\left(\bar{\rho}^{\mathcal{Y}}\right)^{-1}(y)$ (see the examples, $y^A$, $y^B$, $y^C$ and the corresponding rays in $\mathfrak{X}$ in the right panel).
On each ray, the entropy function $\Sigma^{\mathrm{tot}}$ increases when $X$ approaches the origin as shown in the right panel of FIG. \ref{fig:numerical_simulation_sec5}(b). 

\rightline{$\square$}

Finally, we investigate the case $\varphi^{*}\left(y^{\mathrm{E}\mathrm{Q}}\right)-\tilde{\Pi}>0$, in which the growth of the system is realized. 
In this case, a region $\mathfrak{R}^{\mathcal{Y}}(\tilde{\Pi},\tilde{\mu})$ $\subset$ $\mathcal{I}^{\mathcal{Y}}(\tilde{\Pi},\tilde{\mu})$ exists such that $K^{\mathcal{Y}}(y)$ is positive: 
\begin{equation}
    \mathfrak{R}^{\mathcal{Y}}\left(\tilde{\Pi},\tilde{\mu}\right):=\left\{ y|y\in\mathcal{I}^{\mathcal{Y}}\left(\tilde{\Pi},\tilde{\mu}\right),K^{\mathcal{Y}}(y)>0 \right\}.\label{regionY}
\end{equation}
Also, by taking into account the definitions of $\mathcal{I}^{\mathcal{Y}}(\tilde{\Pi},\tilde{\mu})$ and $K^{\mathcal{Y}}(y)$, given in Eqs. (\ref{ibm_Y}) and (\ref{Kappa_Y}), this region can be represented by the intersection: 
\begin{equation}
    \mathfrak{R}^{\mathcal{Y}}\left(\tilde{\Pi},\tilde{\mu}\right)=\mathcal{I}^{\mathcal{Y}}\left(\tilde{\Pi},\tilde{\mu}\right)\cap\mathcal{Z}^\mathcal{Y}\left(\tilde{\mu}\right),\label{IintZ}
\end{equation}
where $\mathcal{Z}^\mathcal{Y}(\tilde{\mu})$ is the larger region: 
\begin{equation}
    \mathcal{Z}^\mathcal{Y}\left(\tilde{\mu}\right):=\left\{y|\varphi^{*}\left(y^{\mathrm{E}\mathrm{Q}}\right)-\varphi^{*}\left(y\right)-\mathcal{D}^{\mathcal{Y}}\left[y^{\mathrm{E}\mathrm{Q}}||y \right]>0\right\}.\label{DEFZ}
\end{equation}
The existence of $\mathfrak{R}^{\mathcal{Y}}(\tilde{\Pi},\tilde{\mu})$ is proved in Appendix D.
Consequently, a ray $\left(\bar{\rho}^{\mathcal{Y}}\right)^{-1}(y)$ for any $y \in \mathfrak{R}^{\mathcal{Y}}(\tilde{\Pi},\tilde{\mu})$ also exists in $\mathfrak{X}$; 
and, on every ray $\left(\bar{\rho}^{\mathcal{Y}}\right)^{-1}(y)$, the value $K^{\mathcal{Y}}(y(X))$ is a positive constant. 
Furthermore, since $\Omega(X)$ is an increasing function on the ray, the entropy function $\Sigma^{\mathrm{t}\mathrm{o}\mathrm{t}}$ increases when $X$ diverges along the ray. 
Accordingly, the entropy function is not bounded above, and the system is growing in this case. 

\textbf{ } 

\textit{Example 6:} Consider the autocatalytic motif shown in FIG. \ref{fig:numerical_outline}(a) and assume that $\varphi^* (y^{\mathrm{EQ}})-\tilde{\Pi}>0$. The region $\mathcal{Z}^\mathcal{Y}\left(\tilde{\mu}\right)$ in $\mathcal{Y}$ is indicated by light pink color in the left panel of FIG. \ref{fig:numerical_simulation_sec5}(c). 
Then, the region $\mathfrak{R}^{\mathcal{Y}}(\tilde{\Pi},\tilde{\mu})$ is given by the intersection between the region $\mathcal{Z}^\mathcal{Y}\left(\tilde{\mu}\right)$ and the level hypersurface (the isobaric manifold) $\mathcal{I}^{\mathcal{Y}}(\tilde{\Pi},\tilde{\mu})$.  
For any $y \in \mathfrak{R}^{\mathcal{Y}}(\tilde{\Pi},\tilde{\mu})$, the value $K^{\mathcal{Y}}(y(X))$ is a positive constant. 
Thus, on a ray $\left(\bar{\rho}^{\mathcal{Y}}\right)^{-1}(y)$ in $\mathfrak{X}$ for every $y \in \mathfrak{R}^{\mathcal{Y}}(\tilde{\Pi},\tilde{\mu})$, the entropy function $\Sigma^{\mathrm{t}\mathrm{o}\mathrm{t}}$ increases when $X$ diverges along the ray. 

The region $\mathcal{Z}^\mathcal{Y}\left(\tilde{\mu}\right)$ exists irrespective of the sign of  
$\varphi^*(y^{\mathrm{EQ}}) - \tilde{\Pi}$ as in FIG. \ref{fig:numerical_simulation_sec5}(a, b).
However, in the cases $\varphi^*(y^{\mathrm{EQ}}) - \tilde{\Pi} \leq 0$, the intersection with the isobaric manifold $\mathcal{I}^{\mathcal{Y}}(\tilde{\Pi},\tilde{\mu})$ does not exist. 

\rightline{$\square$}

The above three situations are summarized as follows:
\begin{thm}
If and only if the reservoir condition satisfies $\varphi^{*}\left( y^{\mathrm{E}\mathrm{Q}} \right)-\tilde{\Pi} = 0$, where $y^{\mathrm{E}\mathrm{Q}} = -\tilde{\mu}OS^{-1}$, 
equilibrium states exist and the system converges to one of them. 
Furthermore, if and only if $\varphi^{*}\left( y^{\mathrm{E}\mathrm{Q}} \right)-\tilde{\Pi} < 0$, 
the system eventually shrinks and finally vanishes. By contrast, if and only if $\varphi^{*}\left( y^{\mathrm{E}\mathrm{Q}} \right)-\tilde{\Pi} > 0$, the system is growing. 
\label{thm1}
\end{thm}

Based on physical intuition, one expects that the fate of the system is classified by a ``gradient" induced by the intensive variables $(\tilde{\Pi},\tilde{\mu})$ in the reservoir. 
The above theorem makes this intuition precise in the sense that $\varphi^{*}(y^{\mathrm{E}\mathrm{Q}})-\tilde{\Pi}$ plays the role of this gradient.  
In fact, $\varphi^{*}(y^{\mathrm{E}\mathrm{Q}})-\tilde{\Pi}$ is represented by the intensive variables $(\tilde{\Pi},\tilde{\mu})$, because $y^{\mathrm{E}\mathrm{Q}}$ is determined only by the chemical potential $\tilde{\mu}$ in the reservoir through Eq. (\ref{y_EQ}). Furthermore, when the gradient is balanced, i.e., $\varphi^{*}(y^{\mathrm{E}\mathrm{Q}})-\tilde{\Pi} = 0$, the system converges to an equilibrium state. By contrast, when the gradient is not balanced, i.e., $\varphi^{*}(y^{\mathrm{E}\mathrm{Q}})-\tilde{\Pi} \neq 0$, the system never reaches an equilibrium state. 

A more precise explanation of the gradient is as follows.
On the one hand, the chemical reactions in the system aim to achieve the state $y^{\mathrm{E}\mathrm{Q}}$, the pressure at which is $\varphi^*(y^{\mathrm{E}\mathrm{Q}})$. 
On the other hand, the internal pressure $\varphi^{*}(y)$ of the system always balances with $\tilde{\Pi}$, owing to the fast dynamics.
The gradient $\varphi^{*}\left( y^{\mathrm{E}\mathrm{Q}} \right)-\tilde{\Pi}$ represents the difference between them. 
When $\varphi^{*}\left( y^{\mathrm{E}\mathrm{Q}} \right)-\tilde{\Pi} = 0$, 
the target pressure $\varphi^{*}( y^{\mathrm{E}\mathrm{Q}})$ coincides with the reservoir pressure $\tilde{\Pi}$. 
Then, the system converges to an equilibrium state. 
In the case that the target pressure is smaller than $\tilde{\Pi}$ (i.e., $\varphi^{*}\left( y^{\mathrm{E}\mathrm{Q}} \right)-\tilde{\Pi} < 0$), 
the chemical reactions attempt to decrease the internal pressure $\varphi^{*}(y)$ from $\tilde{\Pi}$ in each time step, but the system immediately regains $\varphi^{*}(y)=\tilde{\Pi}$. 
This infinitesimal and instantaneous pressure gap between the system and the reservoir leads to the shrinking and the vanishing of the system. 
By contrast, if the target pressure is larger than $\tilde{\Pi}$ (i.e., $\varphi^{*}\left( y^{\mathrm{E}\mathrm{Q}} \right)-\tilde{\Pi} > 0$), from the same argument, the system eventually grows (expands) in each time step and finally diverges. 

\section{VI. Steady growing state}
\label{SecVI}

In this section, we consider the steady growing state and evaluate the entropy production rate at the state. Since the system is assumed to grow, we focus on the case: $\varphi^{*}\left( y^{\mathrm{E}\mathrm{Q}} \right)-\tilde{\Pi} > 0$. 
The steady growing state $x_{\mathrm{SG}}$ is defined as a state such that the density $x(t)=X(t)/\Omega(t)$ is kept constant in the time evolution and $\dot{\Omega}(t)$ is positive, where the dot denotes the time derivative. At this state, the number of confined chemicals $X(t)$ evolves only on a ray $\bar{\rho}_\mathcal{X}^{-1}(x_{\mathrm{SG}})$, because $X(t)=\Omega(t)x_{\mathrm{SG}}$.   

In order for $x_{\mathrm{SG}}$ to be the steady growing state, the entropy production rate at this state must be positive, $\dot{\Sigma}^{\mathrm{tot}}\left(\Omega(t)x_{\mathrm{SG}}\right) > 0$, and, at the same time, the volume must be increasing, i.e., $\dot{\Omega}(t) > 0$.
By substituting $X(t)=\Omega(t)x_{\mathrm{SG}}$ into Eq. (\ref{ent27}), we get 
\begin{equation}
    \displaystyle \Sigma^{\mathrm{t}\mathrm{o}\mathrm{t}}\left(\Omega\left(t\right)x_{\mathrm{S}\mathrm{G}}\right)=\frac{\Omega\left(t\right)}{\tilde{T}}\left\{y_{i}^{\mathrm{E}\mathrm{Q}}-\partial_{i}\varphi\left(x_{\mathrm{S}\mathrm{G}}\right)\right\}x_{\mathrm{S}\mathrm{G}}^{i},\label{ent31}
\end{equation}
where we use $x(\Omega(t)x_{\mathrm{SG}})=\rho_{\mathcal{X}}\left(\Omega\left(t\right)x_{\mathrm{S}\mathrm{G}}\right)=x_{\mathrm{S}\mathrm{G}}$. 
By rearranging Eq. (\ref{ent31}) as in Eq. (\ref{ent_tot_bd}), we have 
\begin{equation}
    \Sigma^{\mathrm{t}\mathrm{o}\mathrm{t}}\left(\Omega\left(t\right)x_{\mathrm{S}\mathrm{G}}\right)=\frac{\Omega(t)}{\tilde{T}} K^{\mathcal{Y}}\left(y^{\mathrm{SG}}\right),
\end{equation}
where $y^{\mathrm{S}\mathrm{G}}:=\partial\varphi\left(x_{\mathrm{S}\mathrm{G}}\right)$ and $K^{\mathcal{Y}}\left(y^{\mathrm{SG}}\right)$ is defined in Eq. (\ref{Kappa_Y}). 
Since $K^{\mathcal{Y}}\left(y^{\mathrm{SG}}\right)$ is kept constant with time, the entropy production rate can be represented as 
\begin{equation}
\dot{\Sigma}^{\mathrm{tot}}\left(\Omega\left(t\right)x_{\mathrm{S}\mathrm{G}}\right) = \frac{\dot{\Omega}(t)}{\tilde{T}} K^{\mathcal{Y}}\left(y^{\mathrm{SG}}\right)>0.
\end{equation}
Because $\dot{\Omega}(t) > 0$ for the steady growing state, $K^{\mathcal{Y}}\left(y^{\mathrm{SG}}\right)$ must be positive. 
Accordingly, the chemical potential for the confined chemicals at the steady growing state, $y^{\mathrm{S}\mathrm{G}}$, must lie in the region $\mathfrak{R}^{\mathcal{Y}}(\tilde{\Pi},\tilde{\mu}) \subset \mathcal{Y}$ 
(see Eqs. (\ref{regionY}), (\ref{IintZ}) and (\ref{DEFZ})).   

To clarify the region of possible $x_{\mathrm{SG}}$ in the density space $\mathcal{X}$, 
we map the region $\mathfrak{R}^{\mathcal{Y}}(\tilde{\Pi},\tilde{\mu})$ to $\mathcal{X}$.
First, we introduce the Bregman divergence on $\mathcal{X}$:
\begin{equation}
\mathcal{D}^{\mathcal{X}}\left[x||x^{\prime}\right]:=\left\{\varphi\left(x\right)-\varphi\left(x^{\prime}\right)\right\}-\partial_{i}\varphi\left(x^{\prime}\right)\left\{x^{i}-\left(x^{\prime}\right)^{i}\right\}. \label{BDX}
\end{equation}
This divergence is related to the one in $\mathcal{Y}$, Eq. (\ref{BDY}), as  $\mathcal{D}^{\mathcal{Y}}\left[y||y^{\prime}\right]=\mathcal{D}^{\mathcal{X}}\left[\partial\varphi^{*}\left(y^{\prime}\right)||\partial\varphi^{*}\left(y\right)\right]$.
Then, the term $K^{\mathcal{Y}}(y)$ defined by Eq. (\ref{Kappa_Y}) is transformed as 
\begin{equation}
K^{\mathcal{X}}\left(x\right)=\varphi^{*}\left(y^{\mathrm{E}\mathrm{Q}}\right)-\tilde{\Pi}-\mathcal{D}^{\mathcal{X}}\left[x||x_{\mathrm{E}\mathrm{Q}}\right], 
\label{Kappa_X}
\end{equation}
where $x_{\mathrm{E}\mathrm{Q}}:=\partial\varphi^{*}\left(y^{\mathrm{E}\mathrm{Q}}\right)$.
Thus, the region in the density space $\mathcal{X}$ can be represented as 
\begin{eqnarray}
\nonumber \mathfrak{R}^{\mathcal{X}}\left(\tilde{\Pi},\tilde{\mu}\right) &:=& \partial \varphi^{*} \left(\mathfrak{R}^{\mathcal{Y}}\right) \\ &=& \left\{x|x\in \mathcal{I}^{\mathcal{X}}\left(\tilde{\Pi},\tilde{\mu}\right),K^{\mathcal{X}}\left(x\right)>0\right\}.
\label{regionX}
\end{eqnarray}
Rewriting this region as the intersection of two submanifolds as in Eq. (\ref{IintZ}), we obtain
\begin{eqnarray}
\nonumber \mathfrak{R}^{\mathcal{X}}\left(\tilde{\Pi},\tilde{\mu}\right) &:=& \partial \varphi^{*} \left(\mathfrak{R}^{\mathcal{Y}}\right)=\partial\varphi^*\left(\mathcal{I}^\mathcal{Y}\right)\cap\partial\varphi^*\left(\mathcal{Z}^\mathcal{Y}\right)\\
&=&\mathcal{I}^{\mathcal{X}}\left(\tilde{\Pi},\tilde{\mu}\right)\cap\mathcal{Z}^\mathcal{X}\left(\tilde{\mu}\right),\label{IintZinX}
\end{eqnarray}
where the region $\mathcal{Z}^\mathcal{X}(\tilde{\mu})$ in $\mathcal{X}$ is represented as 
\begin{equation}
    \mathcal{Z}^\mathcal{X}\left(\tilde{\mu}\right)=\left\{x|x^i\left\{\partial_i\varphi\left(x_\mathrm{EQ}\right)-\partial_i\varphi\left(x\right)\right\}>0\right\}.\label{DEFZX}
\end{equation}

The argument in this section is summarized by the following theorem: 
\begin{thm}
When $\varphi^{*}\left( y^{\mathrm{E}\mathrm{Q}} \right)-\tilde{\Pi} > 0$ and a steady growing state $x_{\mathrm{SG}}$ exists, 
the state $x_{\mathrm{SG}}$ must lie in the region  $\mathfrak{R}^{\mathcal{X}}(\tilde{\Pi},\tilde{\mu})$.
Then, the entropy production rate at the state $x_{\mathrm{SG}}$ is represented as 
\begin{eqnarray}
\nonumber&&\dot{\Sigma}^{\mathrm{tot}}_{\mathrm{SG}}\left(t\right) = \frac{\dot{\Omega}(t)}{\tilde{T}} K^{\mathcal{X}}\left(x_{\mathrm{SG}}\right)\\
&&=\frac{\dot{\Omega}(t)}{\tilde{T}}\{\varphi^{*}\left( y^{\mathrm{E}\mathrm{Q}} \right)-\tilde{\Pi}\}-\frac{\dot{\Omega}(t)}{\tilde{T}}\mathcal{D}^{\mathcal{X}}\left[x_{\mathrm{SG}}||x_{\mathrm{E}\mathrm{Q}}\right].\label{entSGtot}
\end{eqnarray}
\label{thm2}
\end{thm}

The above theorem only identifies the region of possible steady growing states.
The existence and uniqueness of such states are not guaranteed. In addition, which states would be chosen in this region is not determined.   
These details can be analyzed and determined once we specify the functional form of the reaction flux $J(t)$.
For example, we assume that $J(t)$ of the CRS given in FIG. \ref{fig:numerical_outline}(a) obeys mass action kinetics and observe that the steady growing state exists as in FIG. \ref{fig:numerical_outline}(d). 
However, if the functional form of the kinetic law is different from mass action, the existence of the steady growing state is not guaranteed even in the CRS. 

By rearranging Eq. (\ref{entSGtot}), we obtain 
\begin{equation}
    \tilde{T}\frac{\dot{\Sigma}^{\mathrm{tot}}_{\mathrm{SG}}\left(t\right)}{\dot{\Omega}(t)} =\{\varphi^{*}\left( y^{\mathrm{E}\mathrm{Q}} \right)-\tilde{\Pi}\}-\mathcal{D}^{\mathcal{X}}\left[x_{\mathrm{SG}}||x_{\mathrm{E}\mathrm{Q}}\right].\label{costforVG}
\end{equation}
The left hand side of this expression represents the thermodynamic cost for the volume growth, whereas the right hand side can be interpreted as follows. 
The first term represents the external contribution, which is the gradient $\varphi^{*}\left( y^{\mathrm{E}\mathrm{Q}} \right)-\tilde{\Pi}$ induced by the reservoir.
The second term characterizes the internal contribution, which is the Bregman divergence $\mathcal{D}^{\mathcal{X}}\left[x_{\mathrm{SG}}||x_{\mathrm{E}\mathrm{Q}}\right]$ from the equilibrium state $x_\mathrm{EQ}$ to the steady growing state $x_\mathrm{SG}$. 
It gives the total entropy increment during an isochoric relaxation $x_\mathrm{SG} \rightarrow x_\mathrm{EQ}$ (see Ref. \cite{sughiyama01} for details). 
This fact suggests to interpret the second term as the relaxation contribution by the chemical reactions in the system.  
Moreover, in the right hand side, only the steady growing state $x_\mathrm{SG}$ depends on the reaction flux $J(t)$.
As a future perspective, when one designs the reaction flux $J(t)$ to optimize the thermodynamic cost, the expression, Eq. (\ref{costforVG}), may play an important role.

Furthermore, from Eq. (\ref{entSGtot}), we can evaluate the heat dissipation and the work done by the system in the steady growing state based on the first law of thermodynamics. In Appendix E, we summarize the first law in our framework. In Appendix F, we derive expressions for the heat and the work. 

\textbf{ } 

\begin{figure}
    \centering
    \includegraphics[width=0.5\textwidth, clip]{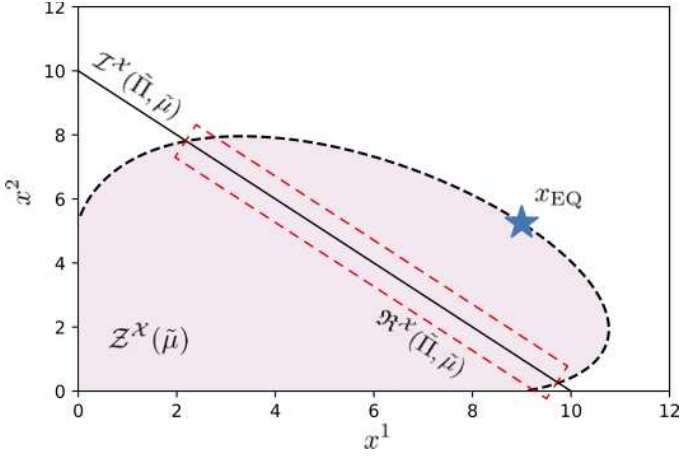}
    \caption{The isobaric manifold $I^{\mathcal{X}}(\tilde{\Pi}, \tilde{\mu})$ in $\mathcal{X}$, corresponding to the case in FIG. \ref{fig:numerical_simulation_sec5}(c). 
    If the system is composed of ideal gas, then $I^{\mathcal{X}}(\tilde{\Pi}, \tilde{\mu})$ is a simplex. The region $\mathcal{Z}^\mathcal{X}(\tilde{\mu})$ is indicated by light pink color. 
    The region $\mathfrak{R}^{\mathcal{X}}(\tilde{\Pi},\tilde{\mu})$ is the intersection between $\mathcal{I}^{\mathcal{X}}(\tilde{\Pi},\tilde{\mu})$ and $\mathcal{Z}^\mathcal{X}(\tilde{\mu})$, which is enclosed by the dashed red rectangle. 
    If a steady growing state exists, it must be in this region. }
    \label{fig:numerical_simulation_sec6}
\end{figure}

\textit{Example 7:} For the example shown in FIG. \ref{fig:numerical_simulation_sec5}(c), in which  $\varphi^* (y^{\mathrm{EQ}})-\tilde{\Pi}>0$ holds, the region $\mathcal{Z}^\mathcal{X}(\tilde{\mu})$ exists in $\mathcal{X}$, as indicated by the light pink color in FIG. \ref{fig:numerical_simulation_sec6}.
Under the ideal gas assumption, the isobaric manifold $\mathcal{I}^{\mathcal{X}}(\tilde{\Pi}, \tilde{\mu})$ is a simplex in $\mathcal{X}$ as we will show in the next section. 
Then, the intersection $\mathfrak{R}^{\mathcal{X}}(\tilde{\Pi},\tilde{\mu})=\mathcal{I}^{\mathcal{X}}(\tilde{\Pi},\tilde{\mu})\cap\mathcal{Z}^{\mathcal{X}}(\tilde{\mu})$ exists as the dashed red rectangle in FIG. \ref{fig:numerical_simulation_sec6}, where $K^{\mathcal{X}}(x)$ is positive for any $x \in \mathfrak{R}^{\mathcal{X}}(\tilde{\Pi},\tilde{\mu})$. 
If a steady growing state $x_{\mathrm{SG}}$ exists, it must be in the region $\mathfrak{R}^{\mathcal{X}}(\tilde{\Pi},\tilde{\mu})$.

\rightline{$\square$}

\section{VII. Ideal gas}
\label{SecVII}
In this section, we demonstrate our framework for CRSs under the ideal gas assumption. To be more precise, we assume that both the system and the reservoir are composed of ideal gas. 

To write down Theorem \ref{thm1} in this situation, we first evaluate the full grand potential density $\varphi^*(y)$. 
The form of the Helmholtz free-energy density for the ideal gas is known as 
\begin{eqnarray}
\displaystyle \nonumber f\left[\tilde{T};n,x\right]&=&n^{m}\displaystyle \mu_{m}^{o}\left(\tilde{T}\right)+R\tilde{T}\sum_{m}\left\{n^{m}\log n^{m}-n^{m}\right\}\\
&&+x^{i}\displaystyle \nu_{i}^{o}\left(\tilde{T}\right)+R\tilde{T}\sum_{i}\left\{x^{i}\log x^{i}-x^{i}\right\},
\end{eqnarray}
where $R$ represents the gas constant; $\mu^{o}(\tilde{T})=\{\mu_{m}^{o}(\tilde{T})\}$ and $\nu^{o}(\tilde{T})=\{\nu_{i}^{o}(\tilde{T})\}$ denote the standard chemical potentials of the open and confined chemicals, respectively. 
Since the partial grand potential density $\varphi[\tilde{T},\tilde{\mu};x]$ can be represented by a variant of the Legendre transformation: 
\begin{equation}
\displaystyle \varphi\left[\tilde{T},\tilde{\mu};x\right]:=\min_{n}\left\{f\left[\tilde{T};n,x\right]-\tilde{\mu}_{m}n^{m}\right\},\label{TPg}
\end{equation}
we get 
\begin{eqnarray}
\displaystyle \nonumber\varphi\left[\tilde{T},\tilde{\mu};x\right]=\varphi\left(x\right)&=&x^{i}\displaystyle \nu_{i}^{o}\left(\tilde{T}\right)+R\tilde{T}\sum_{i}\left\{x^{i}\log x^{i}-x^{i}\right\}\\
&&-R\displaystyle \tilde{T}\sum_{m}e^{\left\{\tilde{\mu}_{m}-\mu_{m}^{o}\left(\tilde{T}\right)\right\}/R\tilde{T}}.\label{IGTPg}
\end{eqnarray}
Also, from the Legendre transformation, Eq. (\ref{LTphi}), the full grand potential density $\varphi^*(y)$ can be expressed as 
\begin{eqnarray}
\displaystyle \nonumber\varphi^{*}\left(y\right)&=&R\displaystyle \tilde{T}\sum_{i}e^{\left\{y_{i}-\nu_{i}^{o}\left(\tilde{T}\right)\right\}/R\tilde{T}}\\
&&+R\displaystyle \tilde{T}\sum_{m}e^{\left\{\tilde{\mu}_{m}-\mu_{m}^{o}\left(\tilde{T}\right)\right\}/R\tilde{T}}.\label{IGDTPg}
\end{eqnarray}
Furthermore, since we have assumed that the reservoir also consists of the ideal gas, 
the chemical potential $\tilde{\mu}$ can be represented as 
\begin{equation}
\tilde{\mu}_{m}=\mu_{m}^{o}\left(\tilde{T}\right)+R\tilde{T}\log\tilde{n}^{m},
\label{IG_chemical_potential}
\end{equation}
where $\tilde{n}=\left\{\tilde{n}^{m}\right\}$ is the density of the open chemicals in the reservoir. 
In addition, for notational simplicity, we define the standard density for the confined chemicals as $x^i_o := e^{ -\nu_i^o(\tilde{T})/R\tilde{T}}$ \cite{n2}. 
Then, Eq. (\ref{IGDTPg}) is rearranged to
\begin{equation}
\displaystyle \varphi^{*}\left(y\right)=R\tilde{T} \sum_{i} x^i_o e^{y_i/R\tilde{T}} + R\tilde{T} \sum_{m} \tilde{n}^m.
\label{IGVP43}
\end{equation}

Next, we calculate the gradient $\varphi^*(y^{\mathrm{EQ}}) - \tilde{\Pi}$ in Theorem \ref{thm1}. 
By defining the standard density for the open chemicals as $n^m_o := e^{ -\mu_m^o(\tilde{T})/R\tilde{T}}$, we get $\tilde{\mu}_m = R \tilde{T} \log \left( \tilde{n}^m/n_o^m \right)$.
Hence, $y^{\mathrm{EQ}} = - \tilde{\mu} O S^{-1}$ in Eq. (\ref{y_EQ}) can be rewritten as
\begin{equation}
    y^{\mathrm{EQ}}_i = R\tilde{T} \log \prod_{m} \left( \frac{n_o^m}{\tilde{n}^m} \right)^{\left( OS^{-1} \right)^m_i}.\label{yEQ55}
\end{equation}
By substituting $y^{\mathrm{EQ}}$ into Eq. (\ref{IGVP43}), we obtain 
\begin{eqnarray}\nonumber
    \varphi^*(y^{\mathrm{EQ}}) - \tilde{\Pi} &=& R\tilde{T} \sum_i \prod_m x^i_o \left( \frac{n_o^m}{\tilde{n}^m} \right)^{\left( OS^{-1} \right)^m_i} \\ && - \left( \tilde{\Pi} - R\tilde{T} \sum_m \tilde{n}^m \right).
    \label{IGD}
\end{eqnarray}
Here, we note that the second line in Eq. (\ref{IGD}) represents the partial pressure that is produced by compositions other than the open chemicals in the reservoir. 
For the ideal gas, Eq. (\ref{IGD}) determines the fate of the system.  

Finally, we specify Theorem \ref{thm2} for the ideal gas. 
The isobaric manifold $\mathcal{I}^{\mathcal{X}} (\tilde{\Pi},\tilde{\mu})$ in Eq. (\ref{ibm_X}) 
is rewritten as 
\begin{equation}
\mathcal{I}^{\mathcal{X}}\left(\tilde{\Pi},\tilde{\mu}\right):=\left\{x| R\tilde{T} \sum_i x^i - \left(\tilde{\Pi} - R\tilde{T} \sum_m \tilde{n}^m \right)=0\right\},
\label{ibm_X_IG}
\end{equation}
which implies the equation of state, $\tilde{\Pi} = R \tilde{T} \left( \sum_i x^i + \sum_m \tilde{n}^m \right)$, and defines a simplex in the density space $\mathcal{X}$.
Also, by using Eq. (\ref{IGTPg}), the region $\mathcal{Z}^\mathcal{X}(\tilde{\mu})$ in Eq. (\ref{DEFZX}) can be expressed as
\begin{equation}
    \mathcal{Z}^\mathcal{X}(\tilde{\mu})=\left\{x| R\tilde{T} \sum_i x^i \log \left( \frac{x^i_{\mathrm{EQ}}}{x^i} \right) > 0\right\},\label{ZX_IG}
\end{equation}
where $x^i_{\mathrm{EQ}}=\partial\varphi^{*}\left(y^{\mathrm{E}\mathrm{Q}}\right)= x_o^i e^{y_i^{\mathrm{EQ}}/R\tilde{T}}$. 
Note that $\sum_ix^i\log(x^i_\mathrm{EQ}/x^i)$ can be negative because $x$ and $x_\mathrm{EQ}$ are not normalized. 
Thus, the region $\mathfrak{R}^{\mathcal{X}}(\tilde{\Pi},\tilde{\mu})$ is given by the intersection between Eqs. (\ref{ibm_X_IG}) and (\ref{ZX_IG}). 
In addition, the Bregman divergence in the density space $\mathcal{X}$, Eq. (\ref{BDX}), reduces to 
the generalized Kullback-Leibler divergence \cite{05,06,07,m1}: 
\begin{equation}
\displaystyle \mathcal{D}^{\mathcal{X}}\left[x||x_{\mathrm{EQ}}\right]=R\tilde{T}\sum_{i}\left[x^{i}\log\frac{x^{i}}{x_{\mathrm{EQ}}^{i}}-\left\{x^{i}-x_{\mathrm{EQ}}^{i}\right\}\right].
\label{GKL}
\end{equation}
Accordingly, the entropy production rate $\dot{\Sigma}^{\mathrm{tot}}_{\mathrm{SG}}(t)$ is evaluated by substituting Eqs. (\ref{IGD}) and (\ref{GKL}) into Eq. (\ref{entSGtot}). 
To obtain the entropy production rate in Eq. (\ref{entSGtot}), we still need to calculate the growth rate $\dot{\Omega}(t)$ and the steady growing state $x_\mathrm{SG}$. 
To compute them, we must determine the functional form of the reaction flux $J(t)$. 
We should recall that Theorem \ref{thm2} only identifies the region of possible steady growing states $x_\mathrm{SG}$.

\textbf{ } 

\textit{Example 8:}
The geometric representations of the examples shown in FIG. \ref{fig:numerical_simulation_sec5} and \ref{fig:numerical_simulation_sec6} are obtained as follows for the ideal gas. 
Before presenting the geometry, we list the given parameters: (1) the stoichiometric matrices $S$ and $O$; (2) the intensive variables $(\tilde{T}, \tilde{\Pi}, \tilde{\mu})$ in the reservoir; 
(3) the standard densities $\{n_o,x_o\}$ or equivalently the standard chemical potentials $\{\mu^o(\tilde{T}),\nu^o(\tilde{T})\}$ for the open and the confined chemicals, 
which are related to each other as $n_o^m = e^{-\mu^o_m(\tilde{T})/R\tilde{T}}$ and $x^i_o = e^{ -\nu_i^o(\tilde{T})/R\tilde{T}}$;
(4) the density $\tilde{n}$ for the open chemicals in the reservoir, which leads to the chemical potential as $\tilde{\mu}_m=\mu^o(\tilde{T})+R\tilde{T}\log\tilde{n}^m$.  

First, we determine the isobaric manifolds $\mathcal{I}^{\mathcal{X}}(\tilde{\Pi},\tilde{\mu})$ and $\mathcal{I}^{\mathcal{Y}}(\tilde{\Pi}, \tilde{\mu})$. 
By using the given $\tilde{T}$, $\tilde{\Pi}$ and $\tilde{n}$, we obtain the isobaric manifold $\mathcal{I}^{\mathcal{X}}(\tilde{\Pi}, \tilde{\mu})$ in the density space $\mathcal{X}$ from Eq. (\ref{ibm_X_IG}) as the simplex in FIG. \ref{fig:numerical_simulation_sec6}.
Also, we can describe the isobaric manifold $\mathcal{I}^{\mathcal{Y}}(\tilde{\Pi}, \tilde{\mu})$ in the chemical potential space $\mathcal{Y}$ by substituting Eq. (\ref{IGVP43}) into Eq. (\ref{ibm_Y}), as shown in the left panels of FIG. \ref{fig:numerical_simulation_sec5}.

Second, we determine the regions $\mathcal{Z}^\mathcal{X}(\tilde{\mu})$ and $\mathcal{Z}^\mathcal{Y}(\tilde{\mu})$. 
By employing Eq. (\ref{yEQ55}), we can calculate $y^\mathrm{EQ}$; and by applying the map $\partial^i \varphi^*(y) = x^i_o e^{y_i/R\tilde{T}}$ to $y^\mathrm{EQ}$, we get $x_{\mathrm{EQ}}$. 
The substitution of $x_\mathrm{EQ}$ into Eq. (\ref{ZX_IG}) leads to $\mathcal{Z}^\mathcal{X}(\tilde{\mu})$ (see the light pink region in FIG. \ref{fig:numerical_simulation_sec6}). 
We also obtain $\mathcal{Z}^\mathcal{Y}(\tilde{\mu})$ by substituting Eqs. (\ref{IGVP43}) and (\ref{BDY}) into Eq. (\ref{DEFZ}) (see the light pink regions in the left panels of FIG. \ref{fig:numerical_simulation_sec5}).

Third, we determine the region $\mathfrak{R}^{\mathcal{X}}(\tilde{\Pi},\tilde{\mu})$ for possible steady growing states $x_\mathrm{SG}$ by Eq. (\ref{IintZinX}).
It is given by the intersection between $\mathcal{I}^{\mathcal{X}}(\tilde{\Pi},\tilde{\mu})$ and $\mathcal{Z}^\mathcal{X}(\tilde{\mu})$, i.e., the dashed red rectangle in FIG. \ref{fig:numerical_simulation_sec6}.

Finally, the entropy function $\Sigma^{\mathrm{tot}}(X)$ on $\mathfrak{X}$ is calculated from Eq. (\ref{ent_tot_bd}). Here, the volume $\Omega(X)$ is obtained from Eq. (\ref{Ophi}) with Eq. (\ref{IGTPg}), i.e., from the equation of state: 
\begin{equation}
\displaystyle
    \Omega(X) = \frac{R\tilde{T} \sum_i X^i}{\tilde{\Pi}-R\tilde{T} \sum_m \tilde{n}^m}.\label{IG_ES}
\end{equation}
Also, the chemical potential (i.e., the map $\rho^{\mathcal{Y}}(X)$) can be calculated as 
\begin{equation}
    y_i(X)=\rho^{\mathcal{Y}}_i(X) = \partial_i \varphi \circ \rho_{\mathcal{X}}(X) = R\tilde{T} \log \left(\frac{X^i}{\Omega(X)x^i_o} \right).
\end{equation}
The heat maps of the right panels of FIG. \ref{fig:numerical_simulation_sec5} are plotted using these equations.

\rightline{$\square$}

\section{VIII. Numerical verification}
\label{SecVIII}
To numerically verify our theory, we deal with the minimal motif of autocatalytic cycles as given in Sec. II, where we assume ideal gas conditions and mass action kinetics. 

The chemical equations of the motif have been represented by two reactions $R_1$ and $R_2$ that involve two confined chemicals $A = (A_1, A_2)$ and two open chemicals $B = (B_1, B_2)$:
\begin{align}\nonumber
    R_1:& A_1 + B_1 \rightleftharpoons A_2 + A_2, \\
    \label{TypeI:R2} R_2:& A_2 \rightleftharpoons A_1 + B_2. 
\end{align}
Also, the stoichiometric matrices are 
\begin{align}
    S = 
    \bordermatrix{     & R_1 & R_2 \cr
               A_1 & -1 & 1 \cr
               A_2 & 2 & -1 \cr
            }, 
    O = 
    \bordermatrix{     & R_1 & R_2 \cr
               B_1 & -1 & 0 \cr
               B_2 & 0 & 1
            }.
\end{align}
The regularity of the matrix $S$ is checked as $\det[S]=-1\neq0$.
Denoting the number of $A = (A_1, A_2)$ by $X = (X^1, X^2)$, the reaction dynamics for the confined chemicals is written as
\begin{equation}
    \frac{dX^{i}}{dt}=S_{r}^{i}J^{r}\left(t\right).\label{SDynamics_example_numerical}
\end{equation}
Furthermore, we assume mass action kinetics for the 
reaction flux $J(t)$:
\begin{eqnarray}\nonumber
J^1(t) &=& w_+^1 X^1 \frac{N^1}{\Omega} - w_-^1 X^2 \frac{X^2}{\Omega}, \\
J^2(t) &=& w_+^2 X^2 - w_-^2 X^1 \frac{N^2}{\Omega}, \label{RF_MA}
\end{eqnarray}
where $N = (N^1, N^2)$ denotes the number of $B = (B_1, B_2)$ in the system. 
The rate constants $w^r_+$ and $w^r_-$ satisfy
\begin{eqnarray}
\displaystyle \log\frac{w_{+}^{r}}{w_{-}^{r}}=-\frac{1}{R\tilde{T}}\left\{\nu_{i}^{0}\left(\tilde{T}\right)S_{r}^{i}+\mu_{m}^{o}\left(\tilde{T}\right)O_{r}^{m}\right\},
\label{IG_LDB}
\end{eqnarray}
which is known as the local detailed balance condition \cite{sughiyama01,kobayashi01,07,06,02}. 

To solve Eq. (\ref{SDynamics_example_numerical}), we need to elucidate 
the behavior of $N$ and $\Omega$.
For the ideal gas, the density $N/\Omega$ of the open chemicals in the system coincides with the density $\tilde{n}$ in the reservoir, which is a constant in time (see Appendix G). 
In addition, $\Omega$ is given by the equation of state as Eq. (\ref{IG_ES}).
Thus, Eq. (\ref{RF_MA}) can be rearranged as 
\begin{eqnarray}\nonumber
J^1(t) &=& \hat{w}_+^1 X^1 - \hat{w}_-^1 \frac{(X^2)^2}{\Omega(X)}, \\
J^2(t) &=& \hat{w}_+^2 X^2 - \hat{w}_-^2 X^1, 
\end{eqnarray}
where we absorb the constant densities of the open chemicals, $N^m/\Omega$, into the rate constants as $\hat{w}^r_+$ and $\hat{w}^r_-$.
Then, the local detailed balance condition in Eq. (\ref{IG_LDB}) can be written as 
\begin{eqnarray}
\displaystyle \log\frac{\hat{w}_{+}^{r}}{\hat{w}_{-}^{r}}=-\frac{1}{R\tilde{T}}\left\{\nu_{i}^{0}\left(\tilde{T}\right)S_{r}^{i}+\tilde{\mu}_{m}O_{r}^{m}\right\},
\end{eqnarray}
and, for our specific example, it reduces to 
\begin{eqnarray}
\displaystyle \frac{\hat{w}_{+}^{1}}{\hat{w}_{-}^{1}}=\frac{x^2_o x^2_o \tilde{n}^1}{x^1_o n^1_o}, \hspace{4mm}
\frac{\hat{w}_{+}^{2}}{\hat{w}_{-}^{2}}=\frac{x^1_o n^2_o}{x^2_o \tilde{n}^2}.
\label{IG_LDB_example}
\end{eqnarray}

Also in this case, $y^{\mathrm{EQ}}$ in Eq. (\ref{yEQ55}) is written as 
\begin{eqnarray}
\nonumber    y^{\mathrm{EQ}}_1 &=& R\tilde{T} \log \left( \frac{n_o^1}{\tilde{n}^1} \right)^{-1} \left( \frac{n_o^2}{\tilde{n}^2} \right)^{2}, \\
    y^{\mathrm{EQ}}_2 &=& R\tilde{T} \log \left( \frac{n_o^1}{\tilde{n}^1} \right)^{-1} \left( \frac{n_o^2}{\tilde{n}^2} \right)^{1}.
\end{eqnarray}
Then, the gradient, Eq. (\ref{IGD}), is represented as 
\begin{eqnarray}\nonumber
    &&\varphi^*(y^{\mathrm{EQ}}) - \tilde{\Pi} = R\tilde{T} \left\{ x_o^1 \left( \frac{n_o^1}{\tilde{n}^1} \right)^{-1} \left( \frac{n_o^2}{\tilde{n}^2} \right)^{2} \right. \\ && \left. + x_o^2 \left( \frac{n_o^1}{\tilde{n}^1} \right)^{-1} \left( \frac{n_o^2}{\tilde{n}^2} \right)^{1} \right\} - \left\{ \tilde{\Pi} - R\tilde{T} \left( \tilde{n}^1 + \tilde{n}^2 \right) \right\}.
\end{eqnarray}
By using this expression, we obtain the following results. 

\begin{figure}[h]
\includegraphics[width=0.5\textwidth, clip]{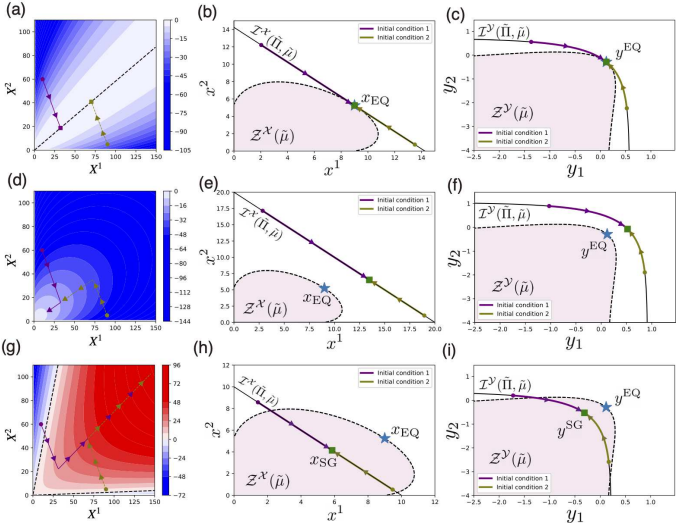}
\caption{Trajectories of the system in the spaces of the number of confined chemicals $\mathfrak{X}$ (left panel; a, d, g), the density space $\mathcal{X}$ (middle panel; b, e, h), and the chemical potential space $\mathcal{Y}$ (right panel; c, f, i) for different pressures $\tilde{\Pi}$ satisfying (top; a,b,c) $\varphi^*(y^{\mathrm{EQ}}) - \tilde{\Pi} = 0$, (middle; d,e,f) $\varphi^*(y^{\mathrm{EQ}}) - \tilde{\Pi} = -5.75$, (bottom; g,h,i) $\varphi^*(y^{\mathrm{EQ}}) - \tilde{\Pi} = 4.25$. 
For our simulation, $x_{\mathrm{EQ}} = (9, 5.25)$ and $y^{\mathrm{EQ}} = (0.118, -0.288)$. 
The parameters of the simulation are fixed as follows: $R = \tilde{T} = 1$, $x^1_o = 8$,  $x^2_o = 7$, $n^1_o = 2$, $\tilde{n}^1=1$, $n^2_o = 3$, $\tilde{n}^2=2$, $\hat{w}^1_- = \hat{w}^2_-=1$. 
The rate constants of forward reactions, $\hat{w}^1_+$ and $\hat{w}^2_+$, are computed by Eq. (\ref{IG_LDB_example}). 
For the initial conditions 1 and 2, we set $(X_1, X_2) = (10, 60)$ and $(90, 5)$, respectively. }
\label{fig:numerical_simulation_sec8}
\end{figure}

In FIG. \ref{fig:numerical_simulation_sec8}, we show the trajectories of the system, from two initial conditions 1 and 2, in the spaces $\mathfrak{X}$, $\mathcal{X}$ and $\mathcal{Y}$.

When the equality $\varphi^*(y^{\mathrm{EQ}}) - \tilde{\Pi} = 0$ holds (see FIG. \ref{fig:numerical_simulation_sec8}(a)), the total entropy function is increasing as the system moves on $\mathfrak{X}$ and converges to a point, denoted by the square, on the equilibrium ray. The point depends on the initial conditions.
In the spaces $\mathcal{X}$ and $\mathcal{Y}$ (FIG. \ref{fig:numerical_simulation_sec8}(b, c)), the system moves on the isobaric manifolds $\mathcal{I}^{\mathcal{X}}(\tilde{\Pi}, \tilde{\mu})$ and $\mathcal{I}^{\mathcal{Y}}(\tilde{\Pi}, \tilde{\mu})$, respectively, and converges to the equilibrium points $x_{\mathrm{EQ}}$ and $y^{\mathrm{EQ}}$, irrespective of the initial conditions.

When $\varphi^*(y^{\mathrm{EQ}}) - \tilde{\Pi} < 0$ (see FIG. \ref{fig:numerical_simulation_sec8}(d)), the system first converges to a ray, and then moves on the ray toward the origin of $\mathfrak{X}$, driven by the increase of the entropy function.
In the spaces $\mathcal{X}$ and $\mathcal{Y}$ (FIG. \ref{fig:numerical_simulation_sec8}(e, f)), the system moves on the isobaric manifolds $\mathcal{I}^{\mathcal{X}}(\tilde{\Pi}, \tilde{\mu})$ and $\mathcal{I}^{\mathcal{Y}}(\tilde{\Pi}, \tilde{\mu})$, respectively, and converges to the points denoted by the squares. These points correspond to the ray on which the system moves toward the origin in $\mathfrak{X}$. 
Therefore, the system finally vanishes. 

Finally, when $\varphi^*(y^{\mathrm{EQ}}) - \tilde{\Pi} > 0$ (FIG. \ref{fig:numerical_simulation_sec8}(g)), the system first converges to a ray, and then moves on the ray away from the origin of $\mathfrak{X}$ with the increase of the entropy function.
In the spaces $\mathcal{X}$ and $\mathcal{Y}$ (FIG. \ref{fig:numerical_simulation_sec8}(h, i)), the system moves on the isobaric manifolds $\mathcal{I}^{\mathcal{X}}(\tilde{\Pi}, \tilde{\mu})$ and $\mathcal{I}^{\mathcal{Y}}(\tilde{\Pi}, \tilde{\mu})$, respectively, and converges to points $x_{\mathrm{SG}}$ and $y^{\mathrm{SG}} = \partial \varphi(x_{\mathrm{SG}})$ denoted by the squares. 
These points correspond to the ray on which the system moves in $\mathfrak{X}$, and are indeed located in $\mathfrak{R}^{\mathcal{X}}(\tilde{\Pi},\tilde{\mu})$ and $\mathfrak{R}^{\mathcal{Y}}(\tilde{\Pi},\tilde{\mu})$ (see also FIG. \ref{fig:numerical_simulation_sec5}(c) and FIG. \ref{fig:numerical_simulation_sec6}).

%\onecolumngrid
%\end{widetext}
%\twocolumngrid

\section{IX. Summary and discussion}
\label{SecIX}

We have established the thermodynamics of growing chemical reaction systems (CRSs) by employing Hessian and projective geometry. 
In this work, we have classified the environmental conditions to distinguish the fate of the CRSs. 
Furthermore, under the growing condition, we have identified the region in the density space where a steady growing state can exist. 
We have also evaluated the entropy production rate in this state. 
It is emphasized again that our results are derived by a general thermodynamic structure without assuming any specific thermodynamic potentials or reaction kinetics; 
i.e., they are obtained based solely on the second law of thermodynamics. 

In this work, we have assumed that the stoichiometric matrix $S$ is regular.
This implies that the system can always relax to the chemical equilibrium state when the volume is fixed, i.e., in the isochoric situation \cite{sughiyama01}. 
In other words, the system never reaches a state that continously produces entropy with constant volume, namely, the conventional nonequilibrium steady state (NESS) \cite{07,06,05,04,m6,m5,m4,m1}. 
Accordingly, the nonequilibrium states treated here, notably the steady growing state, are realized due to the change of the volume. 
This nonequilibrium state with changing volume originates in the extensivity of thermodynamics and should be distinct from the conventional NESS with constant volume.

If the matrix $S$ has a nontrivial right null space ($\dim{\rm Ker}[S] \neq 0$), the system may relax to the NESS even for a constant volume situation.
Such a nongrowing but nonequilibrium state is also biologically relevant, for example, the stationary phase of cells \cite{bergkessel01,koch01,balaban01,maitra01,himeoka02}. 
It is a major challenge for the future to clarify how the nonequilibrium state caused by volume growth and the conventional NESS without growth are compatible and related to each other.

By contrast, if the matrix $S$ has a nontrivial left null space  ($\dim{\rm Ker}[S^T] \neq 0$), the system has conservation laws \cite{sughiyama01,kobayashi01,07,06}.
In our framework, it remains an open problem whether steady growth of the system is possible and realized with the conservation laws.

In this paper, we have assumed the isobaric condition and that the time scale of chemical reactions is the slowest. There may be cases with a  different hierarchy of the time scales, e.g., slow $J_E$, $J_{\Omega}$, and/or $J_D$. Yet, our theoretical framework can still be analogously applied, and how our results change is an important topic for future work.

In our setup, we have ignored the tension of the membrane and assumed that it never bursts (see the caption in FIG. \ref{fig:system}).
However, the membrane does have tension in actual situations.
Even for such cases, our framework can be applied by effectively incorporating the tension into the pressure $\tilde{\Pi}$. 
Furthermore, in biological cells, the membrane molecules themselves are produced and supplied by the intracellular CRS.
In this case, the tension is coupled and changes with the CRS, and therefore the effective $\tilde{\Pi}$ changes with time. 
Accordingly, our theoretical framework needs to be extended further.

Our theory surely serves as the basis of all these extensions, which are important for considering actual and experimental situations of growing protocells or biological cells and also for establishing the physics of self-replicating systems.

\section{Acknowledgement}
The authors thank Kento Nakamura and Genta Chiba for fruitful discussion. This research is supported by JSPS KAKENHI Grant Numbers 19H05799 and 21K21308, and by JST CREST JPMJCR2011 and JPMJCR1927.

\appendix

\section{Appendix A}

In this appendix, we introduce the concept of minimal motifs for growing systems called ``autocatalytic cores". 
It was originally proposed in Ref. \cite{autocatalytic_core} to determine whether a subnetwork embedded in a larger chemical reaction network can be autocatalytic. 
Furthermore, the authors of Ref. \cite{autocatalytic_core} have shown that the regularity of the stoichiometric matrices of the motifs plays an essential role to identify such cores,
by providing the following theorem: 
\begin{thm}
If a chemical reaction network is an autocatalytic core, its stoichiometric matrix $S$ for the confined chemicals must be regular.  
\end{thm}

In the following part, we will briefly review the proof of their theorem (Theorem 3) with our notations.   

First, we mathematically define several conditions for a stoichiometric matrix $S$. 
All of the following definitions are introduced in Ref. \cite{autocatalytic_core}. 
The matrix $S$ is {\it productive}, if ${\rm Im}\left[S\right]\cap \mathbb{R}_{>0}^{\mathcal{N}_{X}}\neq\emptyset$. 
The matrix $S$ is {\it autonomous}, if all column vectors of $S$ contain both strictly negative and strictly positive elements. 
The matrix $S$ is an {\it autocatalytic core}, if $S$ is both productive and autonomous; 
in addition, $S$ satisfies the following condition: if we remove a row or a column vector from $S$, the reduced matrix of $S$ is not both productive and autonomous. 
With this final condition, the matrix $S$ is referred to as minimal, because it does not contain any smaller motifs satisfying both productivity and autonomy. 

Furthermore, we prepare the following terms for the proof: 
If a species is the only reactant of a reaction, we call it the {\it solitary reactant} of the reaction; otherwise, we call it a {\it coreactant} of the reaction. 

The above definitions immediately lead to the following lemmas. 
{\bf (Lemma 1)} We can remove an arbitrary column vector from $S$, while preserving autonomy. 
{\bf (Lemma 2)} We can remove an arbitrary row vector from $S$, while preserving productivity. 
{\bf (Lemma 3)} If a species exists such that it is not the solitary reactant for all reactions in $S$, we can remove the row vector corresponding to the species, while preserving productivity and autonomy. 

With the above definitions and lemmas, we now prove Theorem 3. 
Consider an autocatalytic core $S$ of size $\mathcal{N}_{X}\times \mathcal{N}_{R}$ with rank $\lambda$. 
If we assume $\dim{\rm Ker}\left[S\right]\neq 0$, we can remove a column vector, while preserving ${\rm Im}\left[S\right]$, that is, preserving productivity. 
This contradicts the condition that an autocatalytic core $S$ is minimal. Thus, $\dim{\rm Ker}\left[S\right]$ must be zero, and therefore we have $\lambda=\mathcal{N}_{R}$. 
Furthermore, for every species, some reactions exist such that the species is the solitary reactant of the reactions. Otherwise, because of Lemma 3, we can remove a row vector and this contradicts the condition again that an autocatalytic core $S$ is minimal. Thus, we get $\mathcal{N}_{X}\leq\mathcal{N}_{R}$. 
Since $\lambda\leq \mathcal{N}_{X}, \mathcal{N}_{R}$, it follows that $\lambda=\mathcal{N}_{X}=\mathcal{N}_{R}$. 
This means that $S$ is regular. 

\section{Appendix B}

In this appendix, by employing the second law of thermodynamics, we derive the effective slow dynamics, Eq. (\ref{ReactD}), and the expression of the total entropy function in the slow time scale, Eq. (\ref{enttot_X}). 

Since we have assumed $J_{E}\left(t\right),J_{\Omega}\left(t\right),J_{D}\gg J\left(t\right)$, we can ignore the reaction flux $J\left(t\right)$ in Eqs. (\ref{SDynamics}) and (\ref{RDynamics}) for the fast time scale. 
Then, we get the effective fast dynamics as 
\begin{eqnarray}
\displaystyle \nonumber&&\frac{dE}{dt}=J_{E}\left(t\right),\frac{d\Omega}{dt}=J_{\Omega}\left(t\right),\frac{dN^{m}}{dt}=J_{D}^{m}\left(t\right),\\
&&\displaystyle \frac{d\tilde{E}}{dt}=-J_{E}\left(t\right),\frac{d\tilde{\Omega}}{dt}=-J_{\Omega}\left(t\right),\frac{d\tilde{N}^{m}}{dt}=-J_{D}^{m}\left(t\right).\label{A1}
\end{eqnarray}
The formal solution of Eq. (\ref{A1}) with the initial condition $(E_{0},\Omega_{0},N_{0},\tilde{E}_{0},\tilde{\Omega}_{0},\tilde{N}_{0})$ can be represented as  
\begin{eqnarray}
\nonumber&& E\left(t\right)=E_{0}+\Delta_{E}\left(t\right),\Omega\left(t\right)=\Omega_{0}+\Delta_{\Omega}\left(t\right),\\
\nonumber&& N^{m}\left(t\right)=N_{0}^{m}+\Delta_{N}^{m}\left(t\right),\tilde{E}\left(t\right)=\tilde{E}_{0}-\Delta_{E}\left(t\right),\\
&&\tilde{\Omega}\left(t\right)=\tilde{\Omega}_{0}-\Delta_{\Omega}\left(t\right),\tilde{N}^{m}\left(t\right)=\tilde{N}_{0}^{m}-\Delta_{N}^{m}\left(t\right),\label{Fevo}
\end{eqnarray}
where $\left(\Delta_{E}\left(t\right),\Delta_{\Omega}\left(t\right),\Delta_{N}\left(t\right)\right)$ are the integrals of the flux functions $\left(J_{E}\left(t\right),J_{\Omega}\left(t\right),J_{D}\left(t\right)\right)$ with the initial condition $J_{E}\left(0\right)=J_{\Omega}\left(0\right)=J_{D}\left(0\right)=0$. 
Note that the number of the confined chemicals, $X\left(t\right)$, is a constant in this dynamics. 

By substituting this solution into Eq. (\ref{enttot}), we have the time evolution of the total entropy as 
\begin{eqnarray}
\nonumber\Sigma^{\mathrm{t}\mathrm{o}\mathrm{t}}&&\left(\Delta_{E},\Delta_{\Omega},\Delta_{N}\right)=\Sigma\left[E_{0}+\Delta_{E},\Omega_{0}+\Delta_{\Omega},N_{0}+\Delta_{N},X\right]\\
\nonumber&&+\tilde{\Sigma}_{\tilde{T},\tilde{\Pi},\tilde{\mu}}\left[\tilde{E}_0-\Delta_E,\tilde{\Omega}_0-\Delta_\Omega,\tilde{N}_0-\Delta_N\right]\\
\nonumber=&&\Sigma\left[E_{0}+\Delta_{E},\Omega_{0}+\Delta_{\Omega},N_{0}+\Delta_{N},X\right]\\
&&-\displaystyle \frac{1}{\tilde{T}}\Delta_{E}-\frac{\tilde{\Pi}}{\tilde{T}}\Delta_{\Omega}+\frac{\tilde{\mu}_{m}}{\tilde{T}}\Delta_{N}^{m}+\mathrm{c}\mathrm{o}\mathrm{n}\mathrm{s}\mathrm{t}.,\label{Fenttot}
\end{eqnarray}
where we use the properties of the reservoir; i.e., $\Delta_{E}\left(t\right)\ll\tilde{E}_{0},\ \Delta_{\Omega}\left(t\right)\ll\tilde{\Omega}_{0},\ \Delta_{N}\left(t\right)\ll\tilde{N}_{0}$, and the Taylor expansion for $\tilde{\Sigma}_{\tilde{T},\tilde{\Pi},\tilde{\mu}}$; 
we also use the thermodynamic relations: $\partial\tilde{\Sigma}_{\tilde{T},\tilde{\Pi},\tilde{\mu}}/\partial\tilde{E}=1/\tilde{T},\ \partial\tilde{\Sigma}_{\tilde{T},\tilde{\Pi},\tilde{\mu}}/\partial\tilde{\Omega}=\tilde{\Pi}/\tilde{T}$ and $\partial\tilde{\Sigma}_{\tilde{T},\tilde{\Pi},\tilde{\mu}}/\partial\tilde{N}^{m}=-\tilde{\mu}_{m}/\tilde{T}$. 
In addition, we abbreviate the constant term $\tilde{\Sigma}_{\tilde{T},\tilde{\Pi},\tilde{\mu}}[\tilde{E}_{0},\tilde{\Omega}_{0},\tilde{N}_{0}]$ to ``$\mathrm{c}\mathrm{o}\mathrm{n}\mathrm{s}\mathrm{t}.$". 
According to the second law, the system must climb up the landscape defined by the concave function $\Sigma^{\mathrm{t}\mathrm{o}\mathrm{t}}\left(\Delta_{E},\Delta_{\Omega},\Delta_{N}\right)$ in the time evolution, and finally converge to its maximum, which is called the equilibrium state. 
Hence, we get 
\begin{eqnarray}
\nonumber\left(\Delta_{E},\Delta_{\Omega},\Delta_{N}\right)&\rightarrow&\left(\left(\Delta_{E}\right)_{\mathrm{Q}\mathrm{E}\mathrm{Q}},\left(\Delta_{\Omega}\right)_{\mathrm{Q}\mathrm{E}\mathrm{Q}},\left(\Delta_{N}\right)_{\mathrm{Q}\mathrm{E}\mathrm{Q}}\right)\\
\displaystyle \nonumber&=&\displaystyle \arg\max_{\Delta_{E},\Delta_{\Omega},\Delta_{N}}\Sigma^{\mathrm{t}\mathrm{o}\mathrm{t}}\left(\Delta_{E},\Delta_{\Omega},\Delta_{N}\right),\\
\end{eqnarray}
where $\left(\cdot\right)_{\mathrm{Q}\mathrm{E}\mathrm{Q}}$ represents the value at the equilibrium state of the fast dynamics. 
However, we call this the quasi-equilibrium state, because we later consider the slow dynamics. 
By using the argument shift $E=E_{0}+\Delta_{E},\ \Omega=\Omega_{0}+\Delta_{\Omega},\ N=N_{0}+\Delta_{N}$ and taking Eq. (\ref{Fenttot}) into account, we get the extensive variables at the quasi-equilibrium state as
\begin{eqnarray}
\nonumber&&\left(E_{\mathrm{Q}\mathrm{E}\mathrm{Q}},\Omega_{\mathrm{Q}\mathrm{E}\mathrm{Q}},N_{\mathrm{Q}\mathrm{E}\mathrm{Q}}\right)\\
\displaystyle \nonumber&&=\arg\max_{E,\Omega,N}\left\{\Sigma\left[E,\Omega,N,X\right]-\frac{1}{\tilde{T}}E-\frac{\tilde{\Pi}}{\tilde{T}}\Omega+\frac{\tilde{\mu}_{m}}{\tilde{T}}N^{m}\right\}.\\\label{VPEN}
\end{eqnarray}

The above characterization of the quasi-equilibrium state by the variational form, Eq. (\ref{VPEN}), can be rearranged by introducing thermodynamic potentials as follows. 
First, we define the Helmholtz free energy as 
\begin{equation}
F\displaystyle \left[\tilde{T};\Omega,N,X\right]:=\min_{E}\left\{E-\tilde{T}\Sigma\left[E,\Omega,N,X\right]\right\}.\label{defF}
\end{equation}
Second, by using the Helmholtz free energy, we introduce the partial grand potential: 
\begin{equation}
\Phi\left[\tilde{T},\tilde{\mu};\Omega,X\right]:=\displaystyle \min_{N}\left\{F\left[\tilde{T};\Omega,N,X\right]-\tilde{\mu}_{m}N^{m}\right\}.\label{defGP}
\end{equation}
With the above two thermodynamic potentials, we can reformulate the variational form, Eq. (\ref{VPEN}), as 
\begin{equation}
\displaystyle \Omega_{\mathrm{Q}\mathrm{E}\mathrm{Q}}\left(X\right)=\arg\min_{\Omega}\left\{\Phi\left[\tilde{T},\tilde{\mu};\Omega,X\right]+\tilde{\Pi}\Omega\right\}.
\end{equation}
The other two values, $E_{\mathrm{Q}\mathrm{E}\mathrm{Q}}$ and $N_{\mathrm{Q}\mathrm{E}\mathrm{Q}}$, can be computed as follows. 
Since the equality, 
\begin{eqnarray}
\nonumber\Phi\left[\tilde{T},\tilde{\mu};\Omega_{\mathrm{QEQ}},X\right]&=&-\tilde{T}\Sigma\left[E_{\mathrm{Q}\mathrm{E}\mathrm{Q}},\Omega_{\mathrm{QEQ}},N_{\mathrm{Q}\mathrm{E}\mathrm{Q}},X\right]\\
&&+E_{\mathrm{Q}\mathrm{E}\mathrm{Q}}-\tilde{\mu}_{m}N_{\mathrm{Q}\mathrm{E}\mathrm{Q}}^{m},\label{A2}
\end{eqnarray}
holds, the partial differentiations of $\Phi[\tilde{T},\tilde{\mu};\Omega_{\mathrm{QEQ}},X]$ with respect to $\tilde{T}$ and $\tilde{\mu}$ lead to 
\begin{eqnarray}
\displaystyle \nonumber\Sigma_{\mathrm{Q}\mathrm{E}\mathrm{Q}}\left(X\right)&=&-\displaystyle \frac{\partial\Phi\left[\tilde{T},\tilde{\mu};\Omega_{\mathrm{Q}\mathrm{E}\mathrm{Q}},X\right]}{\partial\tilde{T}},\\
N_{\mathrm{Q}\mathrm{E}\mathrm{Q}}^{m}\displaystyle \left(X\right)&=&-\displaystyle \frac{\partial\Phi\left[\tilde{T},\tilde{\mu};\Omega_{\mathrm{Q}\mathrm{E}\mathrm{Q}},X\right]}{\partial\tilde{\mu}_{m}}.\label{A3}
\end{eqnarray} 
Here, we used the fact that the implicit differentiations of Eq. (\ref{A2}) with respect to $E_{\mathrm{Q}\mathrm{E}\mathrm{Q}}$ and $N_{\mathrm{Q}\mathrm{E}\mathrm{Q}}$ vanish, due to the critical equations for the variational forms, Eqs. (\ref{defF}) and (\ref{defGP}). 
Note that we did not perform the implicit differentiation with respect to $\Omega_{\mathrm{QEQ}}$, despite it being a function of $(\tilde{T},\tilde{\mu},\tilde{\Pi};X)$. 
Also we denote $\Sigma_{\mathrm{Q}\mathrm{E}\mathrm{Q}}\left(X\right)=\Sigma\left[E_{\mathrm{Q}\mathrm{E}\mathrm{Q}},\Omega_{\mathrm{Q}\mathrm{E}\mathrm{Q}},N_{\mathrm{Q}\mathrm{E}\mathrm{Q}},X\right]$. 
Finally, by substituting Eq. (\ref{A3}) into Eq. (\ref{A2}), we obtain 
\begin{eqnarray}
\displaystyle \nonumber E_{\mathrm{Q}\mathrm{E}\mathrm{Q}}\left(X\right)&=&\displaystyle \Phi\left[\tilde{T},\tilde{\mu};\Omega_{\mathrm{Q}\mathrm{E}\mathrm{Q}},X\right]-\tilde{T}\frac{\partial\Phi\left[\tilde{T},\tilde{\mu};\Omega_{\mathrm{Q}\mathrm{E}\mathrm{Q}},X\right]}{\partial\tilde{T}}\\
&&-\displaystyle \tilde{\mu}_{m}\frac{\partial\Phi\left[\tilde{T},\tilde{\mu};\Omega_{\mathrm{Q}\mathrm{E}\mathrm{Q}},X\right]}{\partial\tilde{\mu}_{m}}.\label{A4}
\end{eqnarray}

By employing the above results in the fast dynamics, we derive the effective slow dynamics, which is the reaction dynamics. 
The time evolutions of the internal energy $E\left(t\right)$, the volume $\Omega\left(t\right)$ and the number of the open chemicals $N\left(t\right)$ for the reaction dynamics are already solved, by using the time evolution of the confined chemicals $X\left(t\right)$ and Eqs. (\ref{A3}) and (\ref{A4}), as
\begin{equation}
E(t)=E_{\mathrm{Q}\mathrm{E}\mathrm{Q}}\left(X\right),\Omega(t)=\Omega_{\mathrm{Q}\mathrm{E}\mathrm{Q}}\left(X\right),N(t)=N_{\mathrm{Q}\mathrm{E}\mathrm{Q}}\left(X\right).\label{Sen}
\end{equation} 
Substituting these evolutions into Eq. (\ref{SDynamics}) and taking Eq. (\ref{RDynamics}) into account, we obtain the effective slow dynamics as
\begin{eqnarray}
\displaystyle \nonumber&&\frac{dX^{i}}{dt}=S_{r}^{i}J^{r}\left(t\right),\mbox{  }\frac{d\tilde{E}}{dt}=-\frac{dE_{\mathrm{Q}\mathrm{E}\mathrm{Q}}\left(X\right)}{dt},\\
\displaystyle \nonumber&&\frac{d\tilde{\Omega}}{dt}=-\frac{d\Omega_{\mathrm{Q}\mathrm{E}\mathrm{Q}}\left(X\right)}{dt},\mbox{  }\frac{d\tilde{N}^{m}}{dt}=O_{r}^{m}J^{r}\left(t\right)-\frac{dN_{\mathrm{Q}\mathrm{E}\mathrm{Q}}^{m}\left(X\right)}{dt},\\\label{appDya}
\end{eqnarray}
which is Eq. (\ref{ReactD}) in the main text. 
Here, we should note that the initial condition for the reservoir $(\tilde{E}\left(0\right),\tilde{\Omega}\left(0\right),\tilde{N}\left(0\right))$ in the slow time scale is determined by the fast dynamics as follows. 
The slow dynamics starts with the quasi-equilibrium state with $X_0$, which is the initial condition for the confined chemicals. 
Thus, $(\tilde{E}\left(0\right),\tilde{\Omega}\left(0\right),\tilde{N}\left(0\right))$ must be $(\tilde{E}_{\mathrm{QEQ}}(X_0),\tilde{\Omega}_{\mathrm{QEQ}}(X_0),\tilde{N}_{\mathrm{QEQ}}(X_0))$. 
Since 
$\left(\Delta_{E}\right)_{\mathrm{Q}\mathrm{E}\mathrm{Q}}=E_{\mathrm{Q}\mathrm{E}\mathrm{Q}}-E_{0}$, 
$\left(\Delta_{\Omega}\right)_{\mathrm{Q}\mathrm{E}\mathrm{Q}}=\Omega_{\mathrm{Q}\mathrm{E}\mathrm{Q}}-\Omega_{0}$,
and $\left(\Delta_{N}\right)_{\mathrm{Q}\mathrm{E}\mathrm{Q}}=N_{\mathrm{Q}\mathrm{E}\mathrm{Q}}-N_{0}$, 
we get, from Eq. (\ref{Fevo}), 
\begin{eqnarray}
\nonumber\tilde{E}_{\mathrm{Q}\mathrm{E}\mathrm{Q}}\left(X_0\right)&=&\tilde{E}_{0}-\left\{E_{\mathrm{Q}\mathrm{E}\mathrm{Q}}\left(X_0\right)-E_{0}\right\}\\
\nonumber\tilde{\Omega}_{\mathrm{Q}\mathrm{E}\mathrm{Q}}\left(X_0\right)&=&\tilde{\Omega}_{0}-\left\{\Omega_{\mathrm{Q}\mathrm{E}\mathrm{Q}}\left(X_0\right)-\Omega_{0}\right\}\\
\tilde{N}_{\mathrm{Q}\mathrm{E}\mathrm{Q}}^{m}\left(X_0\right)&=&\tilde{N}_{0}^{m}-\left\{N_{\mathrm{Q}\mathrm{E}\mathrm{Q}}^{m}\left(X_0\right)-N_{0}^{m}\right\}.\label{appa11}
\end{eqnarray}

%Before closing this appendix, we comment on the extensive variables in the reservoir at the quasi-equilibrium state, $(\tilde{E}_{\mathrm{QEQ}},\tilde{\Omega}_{\mathrm{QEQ}},\tilde{N}_{\mathrm{QEQ}})$.
%Since 
%$\left(\Delta_{E}\right)_{\mathrm{Q}\mathrm{E}\mathrm{Q}}=E_{\mathrm{Q}\mathrm{E}\mathrm{Q}}-E_{0}$, 
%$\left(\Delta_{\Omega}\right)_{\mathrm{Q}\mathrm{E}\mathrm{Q}}=\Omega_{\mathrm{Q}\mathrm{E}\mathrm{Q}}-\Omega_{0}$,
%and $\left(\Delta_{N}\right)_{\mathrm{Q}\mathrm{E}\mathrm{Q}}=N_{\mathrm{Q}\mathrm{E}\mathrm{Q}}-N_{0}$, 
%we get, from Eq. (\ref{Fevo}), 
%\begin{eqnarray}
%\nonumber\tilde{E}_{\mathrm{Q}\mathrm{E}\mathrm{Q}}\left(X\right)&=&\tilde{E}_{0}-\left\{E_{\mathrm{Q}\mathrm{E}\mathrm{Q}}\left(X\right)-E_{0}\right\}\\
%\nonumber\tilde{\Omega}_{\mathrm{Q}\mathrm{E}\mathrm{Q}}\left(X\right)&=&\tilde{\Omega}_{0}-\left\{\Omega_{\mathrm{Q}\mathrm{E}\mathrm{Q}}\left(X\right)-\Omega_{0}\right\}\\
%\tilde{N}_{\mathrm{Q}\mathrm{E}\mathrm{Q}}^{m}\left(X\right)&=&\tilde{N}_{0}^{m}-\left\{N_{\mathrm{Q}\mathrm{E}\mathrm{Q}}^{m}\left(X\right)-N_{0}^{m}\right\}.\label{appa11}
%\end{eqnarray}
%If we set an initial condition for the confined chemicals to $X_0$, the initial condition for the other variables in Eq. (\ref{appDya}), $(\tilde{E}\left(0\right),\tilde{\Omega}\left(0\right),\tilde{N}\left(0\right))$, is given by Eq. (\ref{appa11}) with $X=X_{0}$.  

Next, we derive the expression of the total entropy function in the slow time scale, Eq. (\ref{enttot_X}). 
By solving Eq. (\ref{appDya}), we have 
\begin{eqnarray}
\nonumber X^{i}\left(t\right)&=&X_{0}^{i}+S_{r}^{i}\Xi^{r}\left(t\right),\\
\nonumber\tilde{E}\left(t\right)&=&\tilde{E}\left(0\right)-E_{\mathrm{Q}\mathrm{E}\mathrm{Q}}\left(X\left(t\right)\right),\\
\nonumber\tilde{\Omega}\left(t\right)&=&\tilde{\Omega}\left(0\right)-\Omega_{\mathrm{Q}\mathrm{E}\mathrm{Q}}\left(X\left(t\right)\right),\\
\tilde{N}^{m}\left(t\right)&=&\tilde{N}^{m}\left(0\right)+O_{r}^{m}\Xi^r\left(t\right)-N_{\mathrm{Q}\mathrm{E}\mathrm{Q}}^{m}\left(X\left(t\right)\right),\label{AppSevoinR}
\end{eqnarray}
where $\Xi\left(t\right)=\left\{\Xi^{r}\left(t\right)\right\}$ is the integration of $J\left(t\right)$ with the initial condition $\Xi\left(0\right)=0$; this is known as the extent of reaction in chemistry. 
The substitution of Eqs. (\ref{Sen}) and (\ref{AppSevoinR}) into Eq. (\ref{enttot}) enables us to represent the total entropy as 
\begin{eqnarray}
\displaystyle \nonumber\Sigma^{\mathrm{t}\mathrm{o}\mathrm{t}}&=&\Sigma_{\mathrm{Q}\mathrm{E}\mathrm{Q}}\left(X\right)+\tilde{\Sigma}_{\tilde{T},\tilde{\Pi},\tilde{\mu}}[\tilde{E}\left(0\right)-E_{\mathrm{Q}\mathrm{E}\mathrm{Q}}\left(X\right),\\
\nonumber&&\tilde{\Omega}\left(0\right)-\Omega_{\mathrm{Q}\mathrm{E}\mathrm{Q}}\left(X\right),\tilde{N}\left(0\right)+O\Xi-N_{\mathrm{Q}\mathrm{E}\mathrm{Q}}\left(X\right)]\\
\nonumber&=&\Sigma_{\mathrm{Q}\mathrm{E}\mathrm{Q}}\left(X\right)-\frac{1}{\tilde{T}}E_{\mathrm{Q}\mathrm{E}\mathrm{Q}}\left(X\right)-\frac{\tilde{\Pi}}{\tilde{T}}\Omega_{\mathrm{Q}\mathrm{E}\mathrm{Q}}\left(X\right)\\
\displaystyle &&-\frac{\tilde{\mu}_{m}}{\tilde{T}}\left\{O^m_r\Xi^r-N_{\mathrm{Q}\mathrm{E}\mathrm{Q}}^m\left(X\right)\right\}+\mathrm{const}.,
%N_{\mathrm{Q}\mathrm{E}\mathrm{Q}}^{m}\left(X\right)-\frac{\tilde{\mu}_{m}}{\tilde{T}}O_{r}^{m}\left(S^{-1}\right)_{i}^{r}X^{i}+\mathrm{const}.\\
%\displaystyle \nonumber&=&-\displaystyle \frac{1}{\tilde{T}}\left\{\Phi\left[\tilde{T},\tilde{\mu};\Omega_{\mathrm{Q}\mathrm{E}\mathrm{Q}},X\right]+\tilde{\Pi}\Omega_{\mathrm{Q}\mathrm{E}\mathrm{Q}}\left(X\rig%ht)-y_{i}^{\mathrm{E}\mathrm{Q}}X^{i}\right\}\\
%&&+\mathrm{const}.,
\label{enttot_X_inR}
\end{eqnarray}
where we again employ the Taylor expansion for $\tilde{\Sigma}_{\tilde{T},\tilde{\Pi},\tilde{\mu}}$ and the thermodynamic relations as in Eq. (\ref{Fenttot}).
By using the partial grand potential, Eq. (\ref{A2}), we get the simple expression: 
\begin{eqnarray}
\nonumber\Sigma^{\mathrm{tot}}&=&-\frac{1}{\tilde{T}}\left\{\Phi\left[\tilde{T},\tilde{\mu};\Omega_{\mathrm{Q}\mathrm{E}\mathrm{Q}},X\right]+\tilde{\Pi}\Omega_{\mathrm{QEQ}}+\tilde{\mu}_{m}O^m_r\Xi^r\right\}\\
&&+\mathrm{const}.\label{revadd1}
\end{eqnarray}

If we use the chemical potential $y^{\mathrm{EQ}}$ at the chemical equilibrium state (see Eq. (\ref{revision1})), the last term in Eq. (\ref{revadd1}) can be rearranged as 
\begin{eqnarray}
\nonumber\Sigma^{\mathrm{tot}}&=&-\frac{1}{\tilde{T}}\left\{\Phi\left[\tilde{T},\tilde{\mu};\Omega_{\mathrm{Q}\mathrm{E}\mathrm{Q}},X\right]+\tilde{\Pi}\Omega_{\mathrm{QEQ}}-y^{\mathrm{EQ}}_iS^i_r\Xi^r\right\}\\
&&+\mathrm{const}.\label{revadd2}
\end{eqnarray}
Since we have $S^i_r\Xi^r=X^i-X^i_0$ from the first equation in Eq. (\ref{AppSevoinR}), the total entropy in the slow time scale can be represented by the function of the number of the confined chemicals $X$:
\begin{eqnarray}
\nonumber\Sigma^{\mathrm{t}\mathrm{o}\mathrm{t}}\left(X\right)&=&-\frac{1}{\tilde{T}}\left\{\Phi\left[\tilde{T},\tilde{\mu};\Omega_{\mathrm{Q}\mathrm{E}\mathrm{Q}},X\right]+\tilde{\Pi}\Omega_{\mathrm{Q}\mathrm{E}\mathrm{Q}}\left(X\right)\right.\\
&&\left.-y_{i}^{\mathrm{E}\mathrm{Q}}X^{i}\right\}+\mathrm{const}.,
%\label{enttot_X}
\end{eqnarray}
which is Eq. (\ref{enttot_X}) in the main text. 
Here, $y^{\mathrm{EQ}}$ is calculated as 
\begin{equation}
y_{i}^{\mathrm{E}\mathrm{Q}}=-\tilde{\mu}_{m}O_{r}^{m}\left(S^{-1}\right)_{i}^{r}.
\end{equation}
Note that, for deriving Eq. (\ref{revadd2}), we used the existence of the equilibrium state $y^{\mathrm{EQ}}$, which is the solution to the simultaneous equations, Eq. (\ref{revision1}) \cite{nr1}.

\section{Appendix C}
In this appendix, we show that the volume $\Omega(X)$ uniquely exists for a given $X$. 

First, we show the existence of $\Omega(X)$. 
In ordinary thermodynamics, it is known that the system always relaxes to an equilibrium state in the isothermal, isobaric and ``isochemical-potential" situation without chemical reactions, which is the fast time scale dynamics in this paper. 
This physical fact is mathematically rephrased by the fact that the variational form, Eq. (\ref{OQEQ}), has a minimum for any pressure $\tilde{\Pi}>0$; equivalently, Eq. (\ref{Ophi}) also has a minimum. It implies that we have employed the following assumption: the range of the derivative function with respect to $\Omega$, $\partial \Phi(\tilde{T},\tilde{\mu};\Omega,X) / \partial\Omega$, is $\mathbb{R}_{<0}$ for any $\tilde{T}$, $\tilde{\mu}$ and $X$. 
%Thus, the variational form, Eq. (\ref{OQEQ}), has a minimum for any pressure $\tilde{\Pi}>0$; equivalently, Eq. (\ref{Ophi}) also has a minimum. 

Next, we prove the uniqueness of $\Omega(X)$. 
The critical equation for the variational form, Eq. (\ref{Ophi}):
\begin{equation}
\displaystyle \Omega\left(X\right)=\arg\min_{\Omega}\left\{\Omega\varphi\left(\frac{X}{\Omega}\right)+\tilde{\Pi}\Omega\right\},\label{OphiAC}
\end{equation}
can be computed as 
\begin{equation}
h(\Omega):=\varphi\left(\frac{X}{\Omega}\right)-\frac{X^{i}}{\Omega}\partial_{i}\varphi\left(\frac{X}{\Omega}\right)+\tilde{\Pi}=0.\label{CCAC}
\end{equation}
Here, $\partial_{i}\varphi\left(X/\Omega\right)=\left.\partial\varphi\left(x\right)/\partial x^{i}\right|_{x=X/\Omega}$ and we have defined the function $h(\Omega)$. 
The differentiation of $h(\Omega)$ is given as 
\begin{equation}
    \frac{dh}{d\Omega}=\Omega^{-3}X^i\left[\partial_i\partial_j\varphi\left(\frac{X}{\Omega}\right)\right]X^j.
\end{equation}
Since $\varphi$ is strictly convex, its Hessian $\partial_i\partial_j\varphi$ is positive definite. Thus, the function $h(\Omega)$ is a strictly increasing function for $\Omega>0$. 
Accordingly, the critical equation, Eq. (\ref{CCAC}) has a unique solution for $\Omega$. 
Therefore, the volume $\Omega(X)$ is uniquely determined by a given $X$. 

\section{Appendix D}
In this appendix, we prove that the intersection
\begin{equation}
     \mathfrak{R}^{\mathcal{Y}}\left(\tilde{\Pi},\tilde{\mu}\right)=\mathcal{I}^{\mathcal{Y}}\left(\tilde{\Pi},\tilde{\mu}\right)\cap\mathcal{Z}^\mathcal{Y}\left(\tilde{\mu}\right)
\end{equation}
exists if and only if $\tilde{\Pi}<\varphi^*(y^\mathrm{EQ})$. Here, the isobaric manifold $\mathcal{I}^{\mathcal{Y}}(\tilde{\Pi},\tilde{\mu})$ as given in Eq. (\ref{ibm_Y}) and the region $\mathcal{Z}^\mathcal{Y}(\tilde{\mu})$ as given in Eq. (\ref{DEFZ}) are
\begin{equation}
\mathcal{I}^{\mathcal{Y}}\left(\tilde{\Pi},\tilde{\mu}\right)=\left\{y|\varphi^{*}\left(y\right)-\tilde{\Pi} = 0\right\},
\label{ibm_Y_add}
\end{equation}
and
\begin{equation}
    \mathcal{Z}^\mathcal{Y}\left(\tilde{\mu}\right)=\left\{y|\varphi^{*}\left(y^{\mathrm{E}\mathrm{Q}}\right)-\varphi^{*}\left(y\right)-\mathcal{D}^{\mathcal{Y}}\left[y^{\mathrm{E}\mathrm{Q}}||y \right]>0\right\}.\label{DEFZ_add}
\end{equation}

If $\tilde{\Pi} \geq \varphi^*(y^\mathrm{EQ})$ holds, then $\varphi^{*}(y^{\mathrm{E}\mathrm{Q}})-\tilde{\Pi}-\mathcal{D}^{\mathcal{Y}}[y^{\mathrm{E}\mathrm{Q}}||y] \leq 0$ because of the Bregman divergence $\mathcal{D}^{\mathcal{Y}}[y^{\mathrm{E}\mathrm{Q}}||y] \geq 0$ for any $y$.
Thus, the intersection $\mathfrak{R}^{\mathcal{Y}}(\tilde{\Pi},\tilde{\mu})$ is empty.

\begin{figure}[h]
\includegraphics[width=0.4\textwidth, clip]{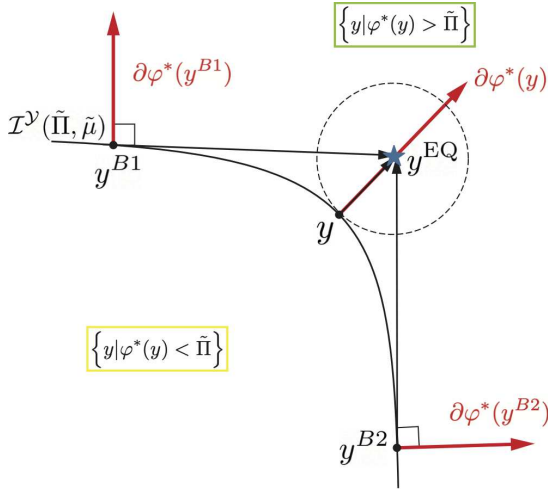}
\caption{Illustration of the proof for the existence of $\mathfrak{R}^{\mathcal{Y}}(\tilde{\Pi},\tilde{\mu})$ for $\tilde{\Pi} < \varphi^*(y^\mathrm{EQ})$.
The solid curve represents the level hypersurface $\{ y | \varphi^*(y) = \tilde{\Pi}\}$, which divides the space $\mathcal{Y}$ into the convex sublevel set (lower left) and the superlevel set (upper right). 
The blue star denotes $y^\mathrm{EQ}$, which is located in the superlevel set. 
The red vectors are the normal vectors $\partial \varphi^{*}(y)$ of the level hypersurface. 
The black vectors are $( y^{\mathrm{EQ}} - y )$. 
The dashed curve expresses the sphere centered at $y^\mathrm{EQ}$. 
By choosing $y$ to be the tangent point between the sphere and the level hypersurface, the inner product between $\partial \varphi^{*}(y)$ and $( y^{\mathrm{EQ}} - y )$ is positive. 
Furthermore, from a similar consideration, the inner product must be positive for any point $y$ between $y^{B_1}$ and $y^{B_2}$.
This region corresponds to $\mathfrak{R}^{\mathcal{Y}}(\tilde{\Pi},\tilde{\mu})$.
%tangent in $\mathcal{I}^{\mathcal{Y}}(\tilde{\Pi},\tilde{\mu})$
%The region $\mathcal{Z}^\mathcal{Y}(\tilde{\mu})$ is shown by light pink color.
%The sublevel set $\{y|\varphi^{*}(y)<\varphi^{*}(y^\mathrm{EQ})\}$, indicated by blue color, is bounded by the level hypersurface  $\mathcal{I}^{\mathcal{Y}}(\varphi^{*}(y^\mathrm{EQ}),\tilde{\mu})$ (solid curve).
%Moreover, the boundary of $\mathcal{Z}^\mathcal{Y}(\tilde{\mu})$ intersects with $\mathcal{I}^{\mathcal{Y}}\left(\varphi^{*}(y^\mathrm{EQ}),\tilde{\mu}\right)$ in exactly one point, namely $y^\textrm{EQ}$.
%If $\tilde{\Pi} \geq \varphi^*(y^\mathrm{EQ})$, the hypersurface $\mathcal{I}^{\mathcal{Y}}(\tilde{\Pi},\tilde{\mu})$ (outer dashed curve) has no intersection with $\mathcal{Z}^\mathcal{Y}(\tilde{\mu})$, 
%whereas for $\tilde{\Pi}' < \varphi^*(y^\mathrm{EQ})$, the hypersurface $\mathcal{I}^{\mathcal{Y}}(\tilde{\Pi}',\tilde{\mu})$ (inner dashed curve) must have the intersection with $\mathcal{Z}^\mathcal{Y}(\tilde{\mu})$.
%The intersection is indicated by the solid black curve inside the curved red rectangle. 
}
\label{fig:appendix}
\end{figure}

If $\tilde{\Pi} < \varphi^*(y^\mathrm{EQ})$, then the intersection $\mathfrak{R}^{\mathcal{Y}}(\tilde{\Pi},\tilde{\mu})$ is not empty by the following argument.
First, we note that the level hypersurface $\mathcal{I}^{\mathcal{Y}}(\tilde{\Pi},\tilde{\mu})=\{ y | \varphi^*(y) = \tilde{\Pi}\}$ divides the space $\mathcal{Y}$ into two regions: one is the sublevel set $\{ y | \varphi^*(y) < \tilde{\Pi}\}$ and the other is the superlevel set $\{ y | \varphi^*(y) > \tilde{\Pi}\}$. 
Because of the convexity of $\varphi^*(y)$, the sublevel set is convex. 
By the assumption $\varphi^*(y^\mathrm{EQ}) > \tilde{\Pi}$, the point $y^\mathrm{EQ}$ lies in the superlevel set (see FIG. \ref{fig:appendix}).
Next, by using the definition of the Bregman divergence, Eq. (\ref{BDY}), the intersection can be rewritten as 
\begin{eqnarray}
    &&\nonumber \mathcal{I}^{\mathcal{Y}}\left(\tilde{\Pi},\tilde{\mu}\right)\cap\mathcal{Z}^\mathcal{Y}\left(\tilde{\mu}\right) \\ &&=  \left\{y|\partial^{i} \varphi^{*}\left(y\right) \left( y^{\mathrm{EQ}}_i - y_i \right) > 0, \varphi^*\left(y\right) = \tilde{\Pi} \right\}.
\end{eqnarray}
The vector $\partial \varphi^{*}(y)$ represents a gradient of the convex function $\varphi^*(y)$, which is a normal vector at $y$ of the level hypersurface. 
Note that the orientation of the normal vector points to the superlevel set (see FIG. \ref{fig:appendix}). 
Also, $( y^{\mathrm{EQ}} - y )$ is a vector from a point $y$ on the level hypersurface to the point $y^{\mathrm{EQ}}$. 
Thus, we can choose $y$ in $\mathcal{I}^{\mathcal{Y}}(\tilde{\Pi},\tilde{\mu})$ such that the inner product between $\partial \varphi^{*}(y)$ and $( y^{\mathrm{EQ}} - y )$ is positive:
Consider the intersection point between the sphere centered at $y^{\mathrm{EQ}}$ which is tangent to the level hypersurface.
This point makes the inner product positive, see FIG. \ref{fig:appendix}. 
This represents $\mathfrak{R}^{\mathcal{Y}}(\tilde{\Pi},\tilde{\mu}) \neq \emptyset$.

%Since the Bregman divergence $\mathcal{D}^{\mathcal{Y}}\left[y^{\mathrm{E}\mathrm{Q}}||y \right]$ in Eq. (\ref{DEFZ}) represents the deviation at the point $y^\mathrm{EQ}$ between the convex function $\varphi^*(y^\mathrm{EQ})$ and tangent plane at the point $y$, 
%a point $y'$ exists in $\mathcal{Y}$ such that $\varphi^{*}\left(y^{\mathrm{E}\mathrm{Q}}\right)-\varphi^{*}\left(y'\right)-\mathcal{D}^{\mathcal{Y}}\left[y^{\mathrm{E}\mathrm{Q}}||y' \right]>0$. 
%Consider the set of all such points, that is the region $\mathcal{Z}^\mathcal{Y}(\tilde{\mu})$, which is unbounded and convex.
%Since $y^\mathrm{EQ}$ is on the boundary of $\mathcal{Z}^\mathcal{Y}(\tilde{\mu})$, the region $\mathcal{Z}^\mathcal{Y}(\tilde{\mu})$ lies inside in the sublevel set $\{y|\varphi^{*}(y)<\varphi^{*}(y^\mathrm{EQ})\}$, which is bounded by $\mathcal{I}^{\mathcal{Y}}(\varphi^{*}(y^\mathrm{EQ}),\tilde{\mu})$ as shown in FIG. \ref{fig:appendix}. 

%If $\tilde{\Pi} \geq \varphi^*(y^\mathrm{EQ})$, then $\mathcal{I}^{\mathcal{Y}}(\tilde{\Pi},\tilde{\mu})$ lies outside the sublevel set $\{y|\varphi^{*}(y)<\varphi^{*}(y^\mathrm{EQ})\}$ and therefore outside of $\mathcal{Z}^\mathcal{Y}(\tilde{\mu})$, i.e. there is no intersection.

%If $\tilde{\Pi} < \varphi^*(y^\mathrm{EQ})$, then $\mathcal{I}^{\mathcal{Y}}(\tilde{\Pi},\tilde{\mu})$ lies inside the sublevel set $\{y|\varphi^{*}(y)<\varphi^{*}(y^\mathrm{EQ})\}$, and due to the unboundedness and convexity of $\mathcal{Z}^\mathcal{Y}(\tilde{\mu})$, the intersection exists.

\section{Appendix E}
In this appendix, we comment on the first law of thermodynamics. 
The internal energy gain should be represented by the heat dissipation $\dot{Q}$ and the work done by the system $\dot{W}$: 
\begin{equation}
\frac{dE_{\mathrm{QEQ}}}{dt}=\frac{d\Phi}{dt}-\tilde{T}\frac{d}{dt}\frac{\partial\Phi}{\partial\tilde{T}}-\tilde{\mu}_m\frac{d}{dt}\frac{\partial\Phi}{\partial\tilde{\mu}_m}=-\dot{Q}-\dot{W},\label{1stlaw}
\end{equation}
where we use Eq. (\ref{A4}) in Appendix B and the dot represents the time derivative. 
Furthermore, for the growing CRSs, the work is composed of the following two kinds: 
\begin{equation}
\dot{W}=\dot{W}^{\mathrm{mech}}+\dot{W}^{\mathrm{chem}},
\end{equation}
where $\dot{W}^{\mathrm{mech}}$ denotes the mechanical work with which the system pushes out the reservoir, due to the growth of the CRSs, and $\dot{W}^{\mathrm{chem}}$ is the work done by the system through the injection of chemicals into the reservoir, which is known as the chemical work. 
These two quantities are given by  
\begin{eqnarray}
\nonumber\dot{W}^{\mathrm{mech}}&:=&\tilde{\Pi}\dot{\Omega}_{\mathrm{QEQ}},\\
\dot{W}^{\mathrm{chem}}&:=&\tilde{\mu}_{m}\frac{d\tilde{N}^{m}}{dt}=\tilde{\mu}_{m}O_{r}^{m}J^{r}-\tilde{\mu}_{m}\frac{dN_{\mathrm{Q}\mathrm{E}\mathrm{Q}}^{m}}{dt},\label{wmc}
\end{eqnarray}
where $d\tilde{N}/dt$ represents the number of the injected chemicals into the reservoir per unit time and we use Eq. (\ref{appDya}). 
From Eqs. (\ref{1stlaw}) and (\ref{wmc}), we can evaluate the heat dissipation as 
\begin{equation}
\dot{Q}=-\frac{dE_{\mathrm{QEQ}}}{dt}-\tilde{\Pi}\dot{\Omega}_{\mathrm{QEQ}}-\tilde{\mu}_{m}O_{r}^{m}J^{r}+\tilde{\mu}_{m}\frac{dN_{\mathrm{Q}\mathrm{E}\mathrm{Q}}^{m}}{dt},
\end{equation}

If we employ the time derivative of the total entropy, Eq. (\ref{enttot_X_inR}):   
\begin{eqnarray}
\nonumber\tilde{T}\dot{\Sigma}^{\mathrm{tot}}&=&\tilde{T}\frac{d\Sigma_{\mathrm{QEQ}}}{dt}-\frac{dE_{\mathrm{QEQ}}}{dt}-\tilde{\Pi}\dot{\Omega}_{\mathrm{QEQ}}\\
&&-\tilde{\mu}_{m}O_{r}^{m}J^{r}+\tilde{\mu}_{m}\frac{dN_{\mathrm{Q}\mathrm{E}\mathrm{Q}}^{m}}{dt},
\end{eqnarray}
we obtain another expression of the heat dissipation: 
\begin{equation}
\dot{Q}=\tilde{T}\dot{\Sigma}^{\mathrm{tot}}-\tilde{T}\frac{d\Sigma_{\mathrm{QEQ}}}{dt}=\tilde{T}\dot{\Sigma}^{\mathrm{tot}}+\tilde{T}\frac{d}{dt}\frac{\partial\Phi}{\partial\tilde{T}},\label{qclau} 
\end{equation}
where we use Eq. (\ref{A3}). 
This expression implies the Clausius inequality: $-\dot{Q}/\tilde{T} \leq d\Sigma_{\mathrm{QEQ}}/dt$, because the total entropy production rate $\dot{\Sigma}^{\mathrm{tot}}$ is nonnegative.
From this expression, we also get another expression of the work as  
\begin{equation}
\dot{W}=-\frac{dE_{\mathrm{QEQ}}}{dt}-\dot{Q}=-\tilde{T}\dot{\Sigma}^{\mathrm{tot}}-\frac{d\Phi}{dt}+\tilde{\mu}_m\frac{d}{dt}\frac{\partial\Phi}{\partial\tilde{\mu}_m},\label{wclau}
\end{equation}
where we use Eq. (\ref{1stlaw}). 

\section{Appendix F}
In this appendix, we evaluate the heat dissipation and the work done by the system in the steady growing state.

From Eqs. (\ref{qclau}) and (\ref{wclau}) in Appendix E and the homogeneity of the partial grand potential $\Phi[\tilde{T},\tilde{\mu};\Omega,X]$, we have 
\begin{eqnarray}
\nonumber\dot{Q}&=&\tilde{T}\dot{\Sigma}^{\mathrm{tot}}+\tilde{T}\frac{d}{dt}\left(\Omega(t)\frac{\partial\varphi(x)}{\partial\tilde{T}}\right),\\
\nonumber\dot{W}&=&-\tilde{T}\dot{\Sigma}^{\mathrm{tot}}-\frac{d\Omega(t)\varphi(x)}{dt}+\tilde{\mu}_m\frac{d}{dt}\left(\Omega(t)\frac{\partial\varphi(x)}{\partial\tilde{\mu}_m}\right),\\
\end{eqnarray}
where we omit the subscript $(\cdot)_{\mathrm{QEQ}}$ for notatinal simplicity as in Sec. IV.
Also, $\varphi(x):=\varphi[\tilde{T},\tilde{\mu};x]$ denotes the partial grand potential density. 
By employing Eq. (\ref{ent41}), from which Eq. (\ref{entSGtot}) follows, the work can be rearranged as
\begin{equation}
\dot{W}=-y^{\mathrm{EQ}}_i\frac{d}{dt}\left(\Omega(t)x^i\right)+\tilde{\Pi}\dot{\Omega}+\tilde{\mu}_m\frac{d}{dt}\left(\Omega(t)\frac{\partial\varphi(x)}{\partial\tilde{\mu}_m}\right).\label{work111}
\end{equation}
Since the second term corresponds to the mechanical work (see Eq. (\ref{wmc})), the chemical work can be represented as 
\begin{equation}
\dot{W}^{\mathrm{chem}}=-y^{\mathrm{EQ}}_i\frac{d}{dt}\left(\Omega(t)x^i\right)+\tilde{\mu}_m\frac{d}{dt}\left(\Omega(t)\frac{\partial\varphi(x)}{\partial\tilde{\mu}_m}\right).\label{work112}
\end{equation}

For the steady growing state $x_{\mathrm{SG}}$, the above equations are further simplified as follows. 
Since $\varphi(x_{\mathrm{SG}})$ is constant with time, we obtain the heat and the work at $x_{\mathrm{SG}}$ as 
\begin{eqnarray}
\nonumber\dot{Q}_{\mathrm{SG}}&=&\tilde{T}\dot{\Sigma}^{\mathrm{tot}}_{\mathrm{SG}}+\dot{\Omega}\tilde{T}\frac{\partial\varphi(x_{\mathrm{SG}})}{\partial\tilde{T}},\\
\dot{W}_{\mathrm{SG}}&=&-\tilde{T}\dot{\Sigma}^{\mathrm{tot}}_{\mathrm{SG}}-\dot{\Omega}\varphi(x_{\mathrm{SG}})+\dot{\Omega}\tilde{\mu}_m\frac{\partial\varphi(x_{\mathrm{SG}})}{\partial\tilde{\mu}_m}.\label{qwsg}
\end{eqnarray}
Also, Eqs. (\ref{work111}) and (\ref{work112}) lead to 
\begin{eqnarray}
\nonumber\dot{W}_{\mathrm{SG}}&=&\dot{\Omega}\left[-y^{\mathrm{EQ}}_i x^i_{\mathrm{SG}}+\tilde{\Pi}+\tilde{\mu}_m\frac{\partial\varphi(x_{\mathrm{SG}})}{\partial\tilde{\mu}_m}\right],\\
\dot{W}^{\mathrm{chem}}_{\mathrm{SG}}&=&\dot{\Omega}\left[-y^{\mathrm{EQ}}_i x^i_{\mathrm{SG}}+\tilde{\mu}_m\frac{\partial\varphi(x_{\mathrm{SG}})}{\partial\tilde{\mu}_m}\right].\label{work114}
\end{eqnarray}
If we can experimentally observe the growth rate $\dot{\Omega}$ and the density profile of the confined chemicals $x_{\mathrm{SG}}$ at the steady growing state, we can evaluate the heat and the work by Eqs. (\ref{qwsg}) and (\ref{work114}).

\section{Appendix G}
In the slow dynamics, the system is always in the quasi-equilibrium state, and therefore 
the number of open chemicals $N(X)$ can be evaluated in Eq. (\ref{secIII:14}) as  
\begin{equation}
    \displaystyle N^{m}\left(X\right)=-\displaystyle \frac{\partial\Phi\left[\tilde{T},\tilde{\mu};\Omega(X),X\right]}{\partial\tilde{\mu}_{m}}.
\end{equation}
Dividing both sides of this equation by $\Omega(X)$ yields
\begin{eqnarray}
\nonumber
n^m(X) &=& \frac{N^m(X)}{\Omega(X)} = \frac{\partial\varphi\left[\tilde{T},\tilde{\mu};X/\Omega(X)\right]}{\partial\tilde{\mu}_m},\\
&=&e^{\left\{\tilde{\mu}_{m}-\mu_{m}^{o}\left(\tilde{T}\right)\right\}/R\tilde{T}} = \tilde{n}^m
\end{eqnarray}
where we use the homogeneity of the partial grand potential $\Phi[\tilde{T},\tilde{\mu};\Omega,X]$.



\begin{thebibliography}{00}
\bibitem{reply2} One may be concerned that the form of the thermodynamic potentials depends on the time scale of the relevant dynamics, e.g., isothermal, isochoric and isobaric processes, etc. However, they are equivalent by a Legendre transformation. In this paper, as we explain in Sec. II, we assume an isobaric process and that the time scale of the chemical reactions is the slowest, but our theoretical framework can be analogously applied to other processes.
\bibitem{nr1} If $S$ is not regular, the existence of the solution to Eq. (\ref{revision1}) is not guaranteed. If the solution does not exist, the system does not have the chemical equilibrium and may relax to a nonequilibrium steady state (see Sec. IX). Furthermore, even if a solution exists, its uniqueness does not hold. This fact implies that the system has multiple equilibrium states and relaxes to one of them depending on the conservation laws \cite{sughiyama01} (see Sec. IX). To avoid unnecessary technicalities and increase the understandability of our theory, we have assumed the regularity of the stoichiometric matrix $S$ in this work.
\bibitem{add_note1} One may concern that the denominator of Eq. (\ref{revision4}) can be negative. However, its positivity is guaranteed by the equation of the state in the reservoir: $\tilde{\Pi} = R\tilde{T} [ \sum_m \tilde{n}^m + \sum_{m'} \tilde{n}_{\mathrm{EX}}^{m'} ]$, where $\tilde{n}_{\mathrm{EX}}$ represents the density of extra  chemicals. Here, the extra chemicals exist only in the reservoir.
\bibitem{n1}  To be more precise, the functional forms of the reaction fluxes $J\left(t\right)$ must be chosen so that $\Sigma^{\mathrm{t}\mathrm{o}\mathrm{t}}\left(X\left(t\right)\right)$ becomes an increasing function with respect to time $t$.
\bibitem{n6} A ray $\mathfrak{r}$ in $\mathfrak{X}$ can be mathematically expressed as follows: for a representative $X_{\mathfrak{r}}\in \mathfrak{X}$, write $\mathfrak{r}=\left\{X|X^{i}=\alpha X_{\mathfrak{r}}^{i},\alpha>0\right\}$.
\bibitem{n7} From Eq. (\ref{map_rho_X}), $\rho_\mathcal{X}\left(X\right)\neq\rho_\mathcal{X}\left(X^{\prime}\right)$ for any two points $X\in \mathfrak{r}$ and $X'\in \mathfrak{r}'$ $ (\mathfrak{r} \neq \mathfrak{r}')$. Thus, the map $\bar{\rho}_\mathcal{X}:\mathcal{P}\mathfrak{X}\rightarrow \mathcal{I}^\mathcal{X}$ is injective.
\bibitem{n2}  The standard density $x_{o}$ satisfies $\partial_i \varphi (x_o)=\nu_o(\tilde{T})+R\tilde{T} \log x^i_0 =0$, which means that $x_o$ is the density that sets the corresponding chemical potential to zero.
\end{thebibliography}
\end{document}